\documentclass[12pt, final]{l4dc2023}

\usepackage{cleveref}
\usepackage{booktabs}
\usepackage{caption}

\newcommand{\eg}{\textit{e.g. }}
\newcommand{\ie}{\textit{i.e. }}
\newcommand{\algo}{{{NeurISS}}}
\newcommand{\xg}{\mathcal{X}^\mathrm{goal}}
\newcommand{\xig}{{\mathcal{X}_i^\mathrm{goal}}}
\newcommand{\xigsample}{\widehat{\mathcal{X}}_i^\mathrm{goal}}
\newcommand{\dist}{\mathrm{dist}}

\newtheorem{theorem}{Theorem}
\newtheorem{lemma}{Lemma}
\newtheorem{proposition}{Proposition}
\newtheorem{definition}{Definition}

\newcommand{\songyuan}[1]{{#1}}

\title[Compositional Neural Certificates for Networked Dynamical Systems]{Compositional Neural Certificates for Networked Dynamical Systems}
\usepackage{times}
\author{%
 \Name{Songyuan Zhang} \Email{szhang21@mit.edu}\\
 \addr Department of Aeronautics and Astronautics, Massachusetts Institute of Technology, Cambridge, MA, USA.
 \AND
  \Name{Yumeng Xiu} \Email{yxiu2@andrew.cmu.edu}\\
 \addr Department of Mechanical Engineering, Carnegie Mellon University, Pittsburgh, PA, USA.
 \AND
 \Name{Guannan Qu} \Email{gqu@andrew.cmu.edu}\\
 \addr Department of Electrical and Computer Engineering, Carnegie Mellon University, Pittsburgh, PA, USA.
 \AND
 \Name{Chuchu Fan} \Email{chuchu@mit.edu}\\
 \addr Department of Aeronautics and Astronautics, Massachusetts Institute of Technology, Cambridge, MA, USA.
}

\begin{document}

\maketitle

\begin{abstract}
Developing stable controllers for large-scale networked dynamical systems is crucial but has long been challenging due to two key obstacles: certifiability and scalability. In this paper, we present a general framework to solve these challenges using compositional neural certificates based on ISS (Input-to-State Stability) Lyapunov functions. Specifically, we treat a large networked dynamical system as an interconnection of smaller subsystems and develop methods that can find each subsystem a decentralized controller and an ISS Lyapunov function; the latter can be collectively composed to prove the global stability of the system. To ensure the scalability of our approach, we develop generalizable and robust ISS Lyapunov functions where a single function can be used across different subsystems and the certificates we produced for small systems can be generalized to be used on large systems with similar structures. We encode both ISS Lyapunov functions and controllers as neural networks and propose a novel training methodology to handle the logic in ISS Lyapunov conditions that encodes the interconnection with neighboring subsystems. We demonstrate our approach in systems including Platoon, Drone formation control, and Power systems. Experimental results show that our framework can reduce the tracking error up to $75\%$ compared with RL algorithms when applied to large-scale networked systems. \footnote{Project website: \href{https://mit-realm.github.io/neuriss-website/}{https://mit-realm.github.io/neuriss-website/}. The appendix can be found on the project website.}
\end{abstract}


\section{Introduction}

Large-scale networked dynamical systems play an important role across a wide spectrum of real-world applications, including power grids \citep{zhao2014design}, vehicle platoons \citep{stankovic2000decentralized}, drone swarms \citep{underactuated}, transportation networks \citep{varaiya2013max}, etc. The control, and in particular stabilization, of such networked systems has long been recognized as a challenging problem as the dimension of the state and input spaces of such networked systems is usually very high, and existing methods often suffer from the ``curse-of-dimensionality"~\citep{powell2007approximate}.

Classical approaches for stabilization of dynamical systems include LQR (Linear Quadratic Regulator) for linear systems \citep{khalil1996robust,dullerud2013course}. For nonlinear systems, certificates like Lyapunov functions can be used to guide the search for a stabilizing controller and certify the stability of the closed-loop system \citep{slotine1991applied}. 
However, certificates are usually constructed on a case-by-case basis. 
\songyuan{While there exist approaches like SOS (Sum-Of-Squares) that can construct certificates for general classes of nonlinear systems \citep{parrilo2000structured}, they are not scalable to large-scale networked dynamical systems, since the number of polynomial coefficients in an SOS program grows exponentially w.r.t. the dimension of system~\citep{parrilo2000structured}.}

A recent line of work parameterizes control certificates (\eg Lyapunov functions, barrier functions) and controllers as neural networks (NNs) and learns them jointly from data \citep{Chang2019,jin2020neural,chow2018lyapunov}. 
\songyuan{They have been successfully applied to nonlinear systems and achieved good performance on complex control tasks \citep{richards2018lyapunov,manek2020learning} thanks to the representation power of NNs.}
However, the existing works mainly focus on single-agent systems with relatively small state space ($\leq10$ dimensions) or multi-agent systems without coupled dynamics~\citep{Chang2019,qin2021learning}. 
Applying these approaches to large-scale networked systems can be challenging due to the exponential growth of the sample complexity and the hardness of training NNs with large input spaces. 
Despite the challenge, many networked dynamics often contain sparse network structures that can be exploited to help training, and the question we try to answer in this paper is: \emph{can we exploit network structure to learn neural certificates and stabilize large scale networked systems in a scalable and effective manner? }

To answer the question, we view the large networked dynamical system as a group of smaller subsystems interconnected through a graph. 
Instead of learning a single certificate for the entire system, we find a decentralized ISS (Input-to-State Stability) Lyapunov function \citep{sontag2013mathematical,liu2011lyapunov,jiang2018small} and a decentralized controller for each subsystem. 
Although ISS Lyapunov functions have been known for decades, it is not straightforward to adapt them as neural certificates for the stabilization of large networked systems due to the following reasons: 1) Existing ISS Lyapunov theory requires checking a condition involving global information of the networked system \citep{liu2011lyapunov} thus is not entirely decentralized; 2) Existing ISS Lyapunov theory requires finding different ISS Lyapunov functions for each subsystem and therefore is computationally expensive for systems with many subsystems; 3) Each subsystem, as well as the corresponding ISS Lyapunov function, are intertwined with neighboring subsystems and therefore cannot be learned straightforwardly as Lyapunov functions for a single system such as that in~\citet {Chang2019}.

To tackle these challenges, we propose \textbf{Neur}al \textbf{ISS} Lyapunov functions (\algo) that make the following \textbf{contributions:} 1) We show that the ISS Lyapunov functions only need to satisfy a \emph{local} condition involving local information from neighboring subsystems in order to collectively constitute a compositional certificate to certify the stability of the entire dynamical system (\Cref{lem:iss_lyapunov}); 
2a) We prove that under certain conditions, the compositional certificate for a small networked system can be generalized to be used on a more extensive system that has a similar structure without re-training (\Cref{lem:generalizaion}), which improves the scalability of the proposed approach as one can reduce a large training task to a smaller training task with a smaller network size; 
2b) We extend the notion of the ISS Lyapunov function to robust ISS Lyapunov function for control-affine systems (\Cref{lem:robust-iss}) so that similar subsystems that have different parameters can share the same ISS Lyapunov functions, which not only reduces the number of ISS Lyapunov functions we need to learn for large-scale networked systems but also improves the robustness of the learned results against model uncertainties. 
3) Furthermore, we develop a novel approach to encode the ISS logic condition that intertwines neighboring subsystems into the training loss function (\Cref{sec:learning}). 

We demonstrate \algo\ using three examples - Power systems, Platoon, and Drone formation control, and show that \algo\ can find certifiably stable controllers for networks of size up to 100 subsystems. Compared with centralized neural certificate approaches, \algo\ reaches similar results in small-scale systems, and can generalize to large-scale systems that centralized approaches cannot scale up to. Compared with LQR, \algo\ can deal with strong coupled networked systems like the microgrids, and reaches smaller tracking errors on both small and large-scale systems. Compared with RL (PPO, LYPPO, MAPPO), our algorithm achieves similar or smaller tracking errors in small systems, and can hugely reduce the tracking errors in large systems (up to $75\%$). 

\textbf{Related Work. } Safe machine learning, neural certificates, and reinforcement learning all have rich literature. Due to space limits, we only mention the most related works. 

\textit{Neural Certificates. } Mostly related is the line of work on learning neural certificates. This line of work focuses on searching for a controller together with a certificate that guarantees the soundness of the controller. Such neural certificates include Lyapunov-like functions for stability guarantees~\citep{Chang2019,jin2020neural,richards2018lyapunov,manek2020learning,abate2020formal,dawson2021charles,gaby2021lyapunov}, barrier functions for safety guarantees~\citep{jin2020neural,qin2021learning,xiao2021barriernet,peruffo2021automated,srinivasan2020synthesis}, contraction metrics for tracking guarantees~\citep{sun2020learning,chou2021model}, etc. Through learning proof of the correctness of the controllers, these approaches address the concerns about the safety, stability, and reliability of the controllers on a large variety of tasks, including precision quadrotor flight through turbulence %induced by propeller wash
~\citep{sun2020learning}, walking under model uncertainties~\citep{castaneda2021gaussian}, tracking with high-dimensional dynamics~\citep{chou2021model}, and safe decentralized control of multi-agent systems~\citep{qin2021learning,meng2021barrier}. Compared to these works, we learn ISS Lyapunov functions, which are decentralized and scalable to large-scale networked systems. We will compare the proposed approach with such neural certificate approaches in \Cref{sec:experiment}.

\textit{ISS Lyapunov Function. } The concept of ISS and ISS Lyapunov function is long established in control theory \citep{sontag2013mathematical}. ISS Lyapunov function for networked dynamical systems was proposed in \citet{jiang1996lyapunov} for a two subsystem case and generalized in \citet{liu2011lyapunov,liu2012lyapunov} for multiple subsystems (cf \citet{jiang2018small} for a review). Compared to these works, our paper builds upon the ISS Lyapunov concept to learn neural certificates for networked systems.  

\textit{Reinforcement Learning (RL). } RL is a popular paradigm in the learning-to-control community with various approaches like (Deep) Q networks \citep{mnih2013playing} , policy optimization \citep{schulman2015trust,schulman2017proximal} and multi-agent versions of them \citep{yu2021surprising} (cf. \citet{sutton2018reinforcement} for a review). However, standard RL is reward-driven and does not formally guarantee stability. Recently, there has been research on learning certificates in the RL process~\citep{berkenkamp2017safe,chow2018lyapunov,cheng2019end,han2020actor,chang2021stabilizing,zhao2021model,qin2021density}, but none of them considers decentralized compositional certificates for networked systems. 
As a result, they are not scalable to large-scale networked systems as they lack the ability to deal with the sheer dimensions of the state space and the exponential growth of sample complexity. We will compare our approach with popular RL algorithms in \Cref{sec:experiment}.

%===============================================================================

\section{Problem Setting}\label{sec:preliminaries}

\songyuan{In this paper, we consider the following networked dynamical system involving $n$ subsystems $\mathcal{N}=\{1,2,\ldots,n\}$. The dynamics of each subsystem are given by
\begin{align}\label{eq:dynamics}
    \dot{x}_i = f_i (x_i,x_{i_1},x_{i_2},\ldots,x_{i_{n_i}},u_i) = f_i(x_i, x_{\mathcal{N}_i},u_i)
\end{align}
where $x_i\in\mathbb{R}^{d_i}$, $u_i\in\mathbb{R}^{p_i}$ is the state and control inputs of each subsystem $i$, and $x_{\mathcal{N}_i}=(x_{i_1},x_{i_2},\ldots,x_{i_{n_i}})$ is used to denote the states of the neighbors of subsystem $i$, $\mathcal{N}_i=\{i_1,i_2,\ldots,i_{n_i}\}$ (not including $i$ itself) which affect the dynamics of the subsystem $i$. }
We use $x = (x_1,\ldots,x_n)$, $u = (u_1,\ldots, u_n)$, and $f = (f_1,\ldots,f_n)$ to denote the vector of states, actions, and dynamics across all subsystems (\ie the overall system), respectively. We also denote $d = \sum_{i=1}^n d_i$ and $p = \sum_{i=1}^n p_i$ to the dimension of $x$ and $u$, respectively.
Our goal is to design \songyuan{a controller $u = \pi(x)$} such that the closed-loop system is asymptotically stable around a goal set $\xg$, formally defined as follows. 

\begin{definition}
Consider a goal set $\xg := \xg_1\times \cdots\times\xg_n$ where each $\xg_i$ is a closed convex subset of $\mathbb{R}^{d_i}$. 
The closed-loop system is globally asymptotically stable \songyuan{about} $\xg_i$ if for any initial state $x(0)$, the trajectory $x(t)$ satisfies $\lim_{t\rightarrow \infty}\dist( x(t) , \xg)= 0$,
where $\dist(x,\xg) := \inf_{y\in \xg} \Vert x - y \Vert$ is the distance between the point $x$ and the set $\xg$. 
\end{definition}

\begin{example} 
\songyuan{A simple networked system is the truck Platoon system with $n+2$ trucks, where the $0$-th (leading) truck can drive freely within the speed and acceleration limits, and the $(n+1)$-th (last) truck will be driven in a way so that the total length of the platoon is roughly kept as a pre-defined constant. We assume other trucks are controllable but can only measure the distance to the two trucks directly in front of and behind themselves. We want to control these trucks so that the trucks in the whole platoon are spread evenly.
Specifically, for truck $i\in\{1,2,...,n\}$, the states are given as $x_i=[p_i^f,p_i^b,v_i]^\top$ and the neighboring subsystems are trucks $\mathcal{N}_i=\{i-1,i+1\}$, where $p_i^f$ is the distance between the $i$-th truck and the $(i-1)$-th truck, $p_i^b$ is the distance between the $i$-th truck and the $(i+1)$-th truck, and $v_i$ is the velocity of the $i$-th truck. The control input is the acceleration of the $i$-th truck. Therefore, the dynamics of truck $i$ is $\dot x_i=[v_{i-1}-v_i,v_i-v_{i+1},a_i]^\top$. 
The goal set for each truck $i$ is uniquely defined as the set of states satisfying $p_i^f=p_i^b$. }
\end{example}

\songyuan{\noindent\textbf{Notations. } Function $\alpha:[0,\infty)\rightarrow[0,\infty)$ is said to be class-$\mathcal{K}$ if $\alpha$ is continuous, strictly increasing, and $\alpha(0) = 0$. Class-$\mathcal{K}$ function $\alpha$ is said to be class-$\mathcal{K}_\infty$ if $\lim_{a\rightarrow+\infty} \alpha(a) = +\infty$. }

%===============================================================================

\section{Compositional Neural Certificates}\label{sec:compositional}
\label{sec:methodology}

\subsection{Decentralized Controller and ISS Lyapunov Functions for Networked Systems}\label{subsec:iss_lyapunov}

\songyuan{Lyapunov functions are widely used to guarantee the stability of dynamical systems (see \Cref{sec:Lyapunov} for an introduction). A common paradigm for stabilizing a dynamical system is to jointly search for a controller $u=\pi(x)$ and a Lyapunov function $V(x)$. 
However, this approach} is not scalable for large-scale networked dynamical systems due to the sheer dimension of the state space, and the controller of the form $u=\pi(x)$, which requires global information of the entire network. 

To address the issues, our framework \algo\ includes two key components: decentralized controllers and compositional certificates. We consider the class of \emph{decentralized controllers} $u_i = \pi_i(x_i)$, which only needs local information within the small subsystem. 
Further, we consider \emph{compositional certificates}, that is, instead of finding a single Lyapunov function $V(x)$ for the whole system, we find one Lyapunov function $V_i(x_i)$ for each subsystem $i$. The individual Lyapunov functions only depend on the subsystem state, which is much smaller in dimension. 
Further, based on \citet{liu2011lyapunov}, we provide the following \Cref{lem:iss_lyapunov} which shows that when the individual Lyapunov functions $V_i$ satisfy an ISS-style condition, they will certify the stability of the entire dynamical system. Since it is the collection of the ISS Lyapunov functions $\{V_i\}_{i=1}^n$ that certify the stability of the entire dynamics, we also call such ISS Lyapunov functions $\{V_i\}_{i=1}^n$ as a ``compositional'' certificate to distinguish them from typical certificates that only contain one Lyapunov function for the entire system. 
A proof of \Cref{lem:iss_lyapunov} is given in \Cref{sec:proofs}. 

\begin{lemma}
\label{lem:iss_lyapunov}
Suppose each subsystem has a decentralized controller $u_i = \pi_i(x_i)$ and a continuously differentiable function $V_i(x_i)$. Suppose: (1) For each $i$, there exists $\mathcal{K}_\infty$ functions $\underline{\alpha}_i,\bar{\alpha}_i$ such that $\underline{\alpha}_i( \dist( x_i,\xig))\leq V_i(x_i)\leq \bar{\alpha}_i(\dist( x_i , \xig))$; (2) For each $i$, there exists $\alpha_i>0$ and class-$\mathcal{K}$ functions $\chi_{ij}, j\in \mathcal{N}_i$ satisfying $\chi_{ij}(a) < a, \forall a>0$, such that $\forall x_i, x_{\mathcal{N}_i}$,
\begin{equation}\label{eq:imply}
\begin{aligned}
&V_i(x_i)\geq \max_{j\in \mathcal{N}_i } \chi_{ij}( V_j (x_j)) 
\quad \Rightarrow \quad [\nabla V_i (x_i) ]^\top f_i(x_i,x_{\mathcal{N}_i}, \pi_i(x_i)) \leq -\alpha_i~ V_i(x_i). 
\end{aligned}
\end{equation} 
Then, the closed-loop system under controllers $\pi_1,\ldots,\pi_n$ is globally asymptotically stable around $\xg$. Such functions $V_i(x_i), i=1,\ldots, n$ are called ISS Lyapunov functions.
\end{lemma}

We note that \Cref{lem:iss_lyapunov} is a variant of the result in \citet{liu2011lyapunov}, in that we explicitly consider the network structure in the dynamics \eqref{eq:dynamics}. As a result, in the ISS implication condition \eqref{eq:imply}, we need to test $V_i(x_i)$ versus the max of $V_j(x_j)$ over only the neighbors $\mathcal{N}_i$, as opposed to the entire network as in \citet{liu2011lyapunov}. This effectively makes \eqref{eq:imply} a condition that can be checked locally at each subsystem. 
One benefit of the local structure in the implication condition \eqref{eq:imply} is that it allows us to use certificates from smaller networks to compose certificates for larger networks that consist of blocks of the smaller networks. We will discuss this in detail in \Cref{subsec:generalizability}. 
Moreover, our results can also be robustified so a single ISS Lyapunov function can be used across different subsystems and handle uncertain parameters in the dynamics. We will explain this in detail in \Cref{subsec:robustness}. 

\subsection{Network Generalizability}\label{subsec:generalizability}
As discussed in \Cref{subsec:iss_lyapunov}, the condition \eqref{eq:imply} only involves $f_i, V_i$, and the Lyapunov functions of neighbors $\{V_j\}_{j\in \mathcal{N}_i}$. With such a local architecture, we present the following \Cref{lem:generalizaion} that shows the decentralized controllers $\pi_i$ and ISS Lyapunov functions $V_i$ for a small system can be ``ported over'' to a larger dynamical system that has a similar symmetric structure to the smaller dynamical system. The proof of \Cref{lem:generalizaion} is postponed to \Cref{sec:proofs}.

\begin{lemma}\label{lem:generalizaion}
Consider a networked dynamical system with node set $\mathcal{N}$, neighborhood sets $\mathcal{N}_i$, and dynamics functions $f_i$, and suppose there exist decentralized controllers $\pi_i$ such that the closed-loop dynamical system admits a compositional certificate $V_i$ that satisfies the conditions in \Cref{lem:iss_lyapunov} with parameters $\chi_{ij}, \alpha_i$. Suppose there is another dynamical system with node set $\tilde{\mathcal{N}}$, neighborhood sets $\tilde{\mathcal{N}}_j$, and dynamics functions $\tilde{f}_i$. Suppose for each $j\in\tilde{\mathcal{N}}$, there exists a one-to-one map $\tau_j: \{j\}\cup \tilde{\mathcal{N}}_j\rightarrow\mathcal{N}$ such that $\tau_j(\tilde{\mathcal{N}}_{j}) = \mathcal{N}_{\tau_j(j)}$, and $\tilde{f}_j = f_{\tau_j(j)}$. Further, suppose $\forall j,j'\in \tilde{\mathcal{N}}$, $\forall\ell \in \tilde{\mathcal{N}}_{j}\cap \tilde{\mathcal{N}}_{j'}$, we have $V_{\tau_j(\ell)} = V_{\tau_{j'}(\ell)}$. Then, $\tilde{\pi}_j = \pi_{\tau_j(j)}$ is a stabilizing controller for the new system with compositional certificate $\tilde{V}_j = V_{\tau_j(j)}$. 
\end{lemma}

\begin{example}
For the Platoon system, using \Cref{lem:generalizaion}, we can prove that a stabilizing controller for a $5$-truck system $\pi_i,i=1,\ldots,5$ can be generalized to any system with $n>5$. Let $
\tilde{\mathcal{N}}$, $\tilde{\mathcal{N}}_i$, $\tilde{f}_i$ be the subsystems, neighborhood, and dynamics functions of the Platoon system with $n$ trucks, and $\mathcal{N}$, $\mathcal{N}_i$, $f_i$ be those of the system with $5$ trucks. Note that in the Platoon system we have $\mathcal{N}_j=\{j-1,j+1\}$ and the same for $\tilde{\mathcal{N}_j}$. We define the one-to-one mapping $\tau_j$ in the following way: For the first truck, let $\tau_1(0)=0,\tau_1(1)=1,\tau_1(2)=2$. For the last truck, let $\tau_n(n-1)=4,\tau_n(n)=5,\tau_n(n+1)=6$. For other trucks $j=2,3,\ldots,n-1$, let $\tau_j(j-1)=2,\tau_j(j)=3,\tau_j(j+1)=4$. Further, let $V_2=V_3=V_4$ and $\pi_2=\pi_3=\pi_4$ in the system with $5$ trucks. In this way, we can check that $\forall j,j'\in \tilde{\mathcal{N}}$, $\forall\ell \in \tilde{\mathcal{N}}_{j}\cap \tilde{\mathcal{N}}_{j'}$, we have $V_{\tau_j(\ell)} = V_{\tau_{j'}(\ell)}$. Then following \Cref{lem:generalizaion}, we can conclude that $\tilde{\pi}_1=\pi_1$, $\tilde{\pi}_n=\pi_5$, $\tilde{\pi}_j=\pi_2=\pi_3=\pi_4,j=2,3,\ldots,n-1$ are stabilizing controllers for the new system with certificates $\tilde{V}_1=V_1,\tilde{V}_n=V_5$, and $\tilde{V}_j=V_2=V_3=V_4,j=2,3,\ldots,n-1$. 
\end{example} 

\subsection{Robust ISS Lyapunov Functions}\label{subsec:robustness}
Many subsystems in a networked system are very similar in terms of dynamics and network structure but may have different parameters. Furthermore, system dynamics may have model uncertainties and unknown parameters~\citep{dawson2021charles}. For instance, in the Platoon example, trucks may have different weights, and the leading truck can have unknown velocity and acceleration, but the platoon system should be stabilized for any driving style of the leading truck.
Having a robust version of ISS Lyapunov functions that can work for a set of different subsystems with different parameters can significantly reduce the number of ISS Lyapunov functions we need to find for a large networked system and also improve the robustness of the resulting controller.
To tackle this, we show that the robust ISS Lyapunov functions can be established for control-affine systems taking the form:
$\dot{x}_i = h_i (x_i,x_{\mathcal{N}_i};\beta) + g_i (x_i,x_{\mathcal{N}_i};\beta)u_i$, 
where $\beta\in\mathcal{B}$ is the parameter of the dynamics that models uncertainties. Such an assumption is not restrictive and can cover a large range of physical systems, \eg, systems following the manipulator function~\citep{underactuated}. Under this assumption, we further introduce robust ISS Lyapunov functions to guarantee the global asymptotic stability of systems with uncertainties. The proof of \Cref{lem:robust-iss} is postponed to \Cref{sec:proofs}. 

\begin{lemma}[\textbf{Robust ISS Lyapunov Functions}]\label{lem:robust-iss}
Given a networked dynamical system with control-affine dynamics with bounded parametric uncertainty $\beta\in\mathcal{B}$, where $\mathcal{B}$ is the convex hull of parameters $\beta_1,\beta_2,\ldots,\beta_{n_\beta}$. If there exists ISS Lyapunov functions $V_i$ satisfying the conditions in \Cref{lem:iss_lyapunov} for each $\beta_j,j\in\{1,2,\ldots,n_\beta\}$, the dynamics $h_i$ and $g_i$ are affine with respect to $\beta$, then the closed-loop system is globally asymptotically stable with any $\beta\in\mathcal{B}$. 
\end{lemma}

\begin{example}
The robust ISS Lyapunov functions can be directly used to the Platoon system. $v_0=v_{n+1}$ is a parameter of this system, which is bounded between $v_\mathrm{min}$ and $v_\mathrm{max}$ by assumption. Following \Cref{lem:robust-iss}, if we can find ISS Lyapunov functions for both $v_\mathrm{min}$ and $v_\mathrm{max}$, we can ensure our system is stable with any velocity of the leading truck. 
\end{example}

%===============================================================================

\section{Learning Compositional Certificates and Controllers } \label{sec:learning}

Based on the compositional certificate developed in \Cref{sec:compositional}, we now focus on jointly learning the individual ISS Lyapunov functions and the decentralized controllers. We note that while there are many existing approaches to learn neural certificates \citep{dawson2021charles,gaby2021lyapunov}, they can not be directly applied here because we have a unique imply condition \eqref{eq:imply} that can not be handled by the existing approaches. We will introduce a novel approach to incorporate the imply condition as specially designed loss terms. To proceed, we start with formally defining the parameterization of the decentralized controllers and the ISS Lyapunov functions. 

\textbf{Controllers and ISS Lyapunov Functions Parameterization.} We focus on decentralized controllers, in which the control $u_i$ of subsystem $i$ only depends on the subsystem state $x_i$, i.e. $u_i = \pi_{i}(x_i;\theta_i)$. Here $\pi_{i}$ is an NN with $\theta_i$ as the parameters. 
\songyuan{We parameterize Lyapunov function $V_i(x_i)$ as $V_i(x_i;S_i,\omega_i,\nu_i) = x_i^\top S_i^\top S_i x_i + p_{i}(x_i;\omega_i)^\top p_{i}(x_i; \omega_i) + q_{i}(x_i;\nu_i)$. }
where $S_i\in\mathbb{R}^{d_i\times d_i}$ is a matrix of parameters, $p_{i}(x_i;\omega_i)$ is an NN with weights $\omega_i$, and $q_{i}(x_i;\nu_i)$ is another NN with weights $\nu_i$ and $\mathrm{ReLU}$ as the output activation function, which is only applied to the output of the NN. The first term in $V_i(x_i)$ is a quadratic term to capture the linear part of the non-linear dynamics. The second term is a sum-of-squares term to capture the polynomial part of the dynamics. Finally, the third term is used to model the residues. Using this form, the ISS Lyapunov function satisfies $V_i(x_i)\geq 0$ by construction. 

\textbf{Sharing ISS Lyapunov Across Subsystems. } Following \Cref{lem:generalizaion} and \Cref{lem:robust-iss}, to reduce the number of neural networks and to make the ISS Lyapunov functions learned in small-scale networked system generalizable to large-scale systems, we use the following weight sharing technique. We let the subsystems share the same ISS Lyapunov functions if their dynamics are similar. 
In addition, we can also let similar subsystems share the same controller. In this way, the number of trainable parameters can be reduced, and we can easily apply the controllers trained in small-scale systems to large-scale systems. For example, in the previous Platoon system, we can let the $j$-th truck, $j=2,3,\ldots,n-1$, share the same ISS Lyapunov function and the same controller. 

\textbf{Gain Function Parameterization. } We use linear functions to model the gain functions $\chi_{ij}$ in \Cref{eq:imply}. 
% To reduce the number of parameters, 
We let $\chi_{ij}(x)=\chi_i(x;k_i)=\mathrm{Sigmoid}(k_i) x,\forall j$, where $k_i$ is a trainable parameter, $x$ is the scalar input of the gain functions, which is always the output of the ISS Lyapunov functions. In this way, the condition $\chi_{ij}(x)<x,\forall x>0$ is satisfied by construction.

\textbf{Loss Functions.} A key challenge in learning ISS Lyapunov functions is how to ensure condition \eqref{eq:imply} is satisfied. We now propose a methodology that promotes \eqref{eq:imply}. Let Boolean $A_i,B_i\in\{0,1\}$ be\footnote{For notational simplicity, from now on we omit all the notations of parameters in the function approximators.} 
\begin{align}
    A_i=V_i(x_i)\geq \max_{j\in \mathcal{N}_i } \chi_{i} (V_j (x_j)), \quad
    B_i
    = [\nabla V_i (x_i) ]^\top f_i(x_i,x_{\mathcal{N}_i},\pi_{i}(x_i)) \leq -\alpha_i V_i(x_i).
\end{align}
Then condition~\eqref{eq:imply} can be written as $A_i\Rightarrow B_i$, which is the same as $\neg A_i \vee B_i$, or $\max\{\neg A_i, B_i\}$. However, this kind of formulation is not trainable for neural networks because Boolean variables are not differentiable. To settle this problem, we introduce the following losses:
\begin{align}
    \mathcal{L}_{A_i}&=\mathrm{ReLU}\left(V_i(x_i)-\max_{j\in \mathcal{N}_i } \chi_{i} (V_j (x_j))+\epsilon_A\right),\label{eq:loss-cond}\\
    \mathcal{L}_{B_i}&=\mathrm{ReLU}\Big([\nabla V_i (x_i) ]^\top f_i(x_i,x_{\mathcal{N}_i},\pi_{i}(x_i))  +\alpha_i V_i(x_i)+\epsilon_B\Big),\label{eq:loss-decent}
\end{align}
where $\epsilon_A$ and $\epsilon_B$ are small parameters that encourages strict satisfactions and generalization abilities~\citep{dawson2021charles}. $\mathcal{L}_{A_i}$ and $\mathcal{L}_{B_i}$ can address the problem introduced by Boolean variables $A_i$ and $B_i$, but they introduce a new problem that $\max\{\neg A_i, B_i\}$ cannot be written as $\max\{\mathcal{L}_{A_i},\mathcal{L}_{B_i}\}$ since the two losses are not comparable. To address this issue, we minimize the loss $\mu_{A_i}\mathcal{L}_{A_i}+\mu_{B_i}\mathcal{L}_{B_i}$ instead, where $\mu_{A_i}$ and $\mu_{B_i}$ are two hyper-parameters for balancing the two losses. 
Note that in practice, since we often simulate the dynamical systems in a discrete way, we can use two ways to calculate $\nabla V_i(x_i)$ in $\mathcal{L}_{B_i}$. 
First, we can directly calculate the gradient of $V_i(x_i)$ w.r.t. $x_i$. For the second method, we can just do a one-step simulation $x_i\rightarrow x_i^\mathrm{next}$, and approximate $[\nabla V_i(x_i)]^\top f_i(x_i,x_{\mathcal{N}_i},\pi_i(x_i))$ with $(V_i(x_i^{\mathrm{next}})-V_i(x_i))/\Delta t$, where $\Delta t$ is the simulation time step. 

\songyuan{To ensure the condition $V_i(x_i)=0$ for $x_i\in\xig$ is satisfied, we introduce another loss term $\frac{1}{| \xigsample |}\sum_{x_i^{\mathrm{goal}} \in \xigsample }|V_i(x_i^{\mathrm{goal}})|$, where $\xigsample$ is a randomly sampled set of states from $\xig$. In addition, we add $\|\pi_{i}(x_i)-u_i^\mathrm{nominal}\|^2$ to the loss, where $u_i^\mathrm{nominal}$ is the control signal calculated by some nominal controller. We use the Droop controller and LQR controller in our experiments. We add the nominal controller so that the learned controller can explore the ``informed region'' near the nominal control signal rather than randomly, in order to accelerate the training. We do not need the nominal controller to be stable or optimal, and the learned controller behaves much better than the nominal controller as shown in \Cref{sec:experiment}. The final loss function used in training is }
\begin{equation}\label{eq:total-loss}
    \begin{aligned}
   & \mathcal{L}= \sum_{i=1}^n \Bigg[\frac{1}{| \xigsample |}\sum_{x_i^{\mathrm{goal}} \in \xigsample }|V_i(x_i^{\mathrm{goal}})|+\mu_{A_i}\mathcal{L}_{A_i}+\mu_{B_i}\mathcal{L}_{B_i}+\mu_\mathrm{ctrl}\|\pi_{i}(x_i)-u_i^\mathrm{nominal}\|^2\Bigg],
    \end{aligned}
\end{equation}
where $\mu_\mathrm{ctrl}$ is a tuning parameter, and the training parameters are $\theta,S,\omega,\nu,k$. 

\textbf{Training Procedure. } During training, we draw samples by randomly sampling states in the state space and in the goal set. We first initialize the controller by minimizing the loss $\|\pi_i(x_{i})-u_i^\mathrm{nominal}\|^2$, and then fix the controller to initialize the ISS Lyapunov function by minimizing the loss $\mathcal{L}_{B_i}$. After the initialization, we minimize loss~\eqref{eq:total-loss} to train the controllers, the ISS Lyapunov functions, and the gain functions jointly. The contour plots of the learned robust ISS Lyapunov functions are provided in Appendix~\ref{sec:add_results}.

%===============================================================================

\section{Experiments}
\label{sec:experiment}

We demonstrate \algo\ in $3$ environments including Power system, Platoon, and Drone, aiming to answer the following questions:  How does \algo\ compare with other algorithms in the case of stabilizing networked systems?  Can \algo\ perform similarly or surpass the centralized controllers in small-scale networked systems? Can \algo\ scale up to large-scale networked systems?
 We provide implementation details, introductions to the systems, and more results in the appendix. 

\songyuan{\textbf{Baselines.} We compare \algo\ with both centralized and decentralized baselines. For centralized ones, we compare with the state-of-the-art RL algorithm PPO \citep{schulman2017proximal}, the RL-with-Lyapunov-critic algorithm LYPPO~\citep{chang2021stabilizing}, and the centralized Neural CLF controller (NCLF)~\citep{dawson2021charles}.
For decentralized ones, we compare with the classical LQR~\citep{kwakernaak1974linear} controller and the multi-agent RL algorithm MAPPO~\citep{yu2021surprising}. 
We hand-craft reward functions based on the common way of designing reward functions for tracking problems for the RL algorithms. 
For LQR, since the agents only have local observations, we calculate the goal point for the LQR controller based on local observations in each time step. 
}

\subsection{Environment Descriptions}
{\textbf{\;\;\;\;\;Power Systems. } We consider two control problems in power systems.  Firstly, }we consider a networked microgrid system introduced in \citet{huang2021neural}, where there is an interconnection of $5$ microgrids.
Each microgrid $i$ has two states $x_i = (\delta_i,E_i)$ where $\delta_i$ is the voltage phase angle and $E_i$ is the voltage magnitude. The goal is to design controllers so that $\delta_i, E_i$ can converge to their reference values $\delta_i^\mathrm{ref},E_i^\mathrm{ref}$. 
{Secondly, we consider a distribution grid voltage control problem \citep{shi2022stability} which we name GridVoltage8. The goal is to drive the distribution grid voltage to the nominal value $1.0$. Due to space limits, more details of the two systems are deferred to \Cref{subsec:environment_detail}.}
\songyuan{Since the dynamics of the power systems are not separable, we use a droop controller as one of the baselines and the nominal controller for \algo\ and NCLF instead of LQR. }

\songyuan{\textbf{Platoon.}
The Platoon system has been introduced in the examples before. We use the LQR controller as the nominal controller of \algo\ and NCLF. Because of the robustness and generalizability of \algo\, we let the controller and the ISS Lyapunov functions of the first and the $n$-th truck share the same weights, while other trucks also share the same weights. We let the controllers of MAPPO share the weights in the same way for a fair comparison. In testing, we let the leading truck's acceleration follow a $\mathrm{sin}$-like curve with clips, which is hard to track. In the small-scale training and testing, we use $n=5$ trucks. In the large-scale testing, we use $n=100$ trucks.}

\textbf{Drone.}
We design the Planar Drone Formation Control environment to further demonstrate the capability of \algo\ in complex networked systems. In this environment, at the beginning of the simulations, the planar drones~\citep{underactuated} stay on the ground. As the simulations start, we want the drones to form a 2D mesh grid while tracking a given trajectory. The states of the drones are modeled as a 2-D platoon system, which is given by $x_i=[p_i^l,p_i^r,p_i^u,p_i^d,\theta_i,v_i^x,v_i^y,\omega_i]^\top$, where $p_i^l,p_i^r,p_i^u,p_i^d$ are the distances from drone $i$ to the left, right, up, down drones, $\theta_i$ is the angle between the drone and the horizontal line, $v_i^x$ and $v_i^y$ are velocities and $\omega_i$ is the angular velocity. The control inputs of each drone are the forces generated by the two propellers. 
\songyuan{We use the LQR controller as the nominal controller for \algo\ and NCLF, and let the controllers in \algo\ and MAPPO share the same weights. Because of the robustness, we also let the ISS Lyapunov functions in \algo\ share the same weights, so we only need $1$ ISS Lyapunov function. In testing, we set the target trajectory to follow a horizontal line with a $\mathrm{sin}$-like acceleration. In the small-scale training and testing, we use $2\times2=4$ drones. In large-scale testing, we use $10\times10=100$ drones. }

\begin{table*}[t]
    \small
    \centering
    \begin{tabular}{c|cccc}
        \toprule
        Environment & Microgrid5 & GridVoltage8 & Platoon5 & Drone2x2 \\
        \midrule
        \algo  & $2027.25\pm18.15$ & $\mathbf{2483.70}\pm1.68$ & $\mathbf{2054.89}\pm95.84$ & $\mathbf{1713.73}\pm13.35$\\
        LQR & ------ & ------ & $1894.64\pm5.20$ & $1209.15\pm5.27$\\
        Droop  & $1431.30\pm55.03$ & $2251.70\pm0.00$ & ------ & ------\\
        PPO  & $1970.89\pm43.97$ & $1086.72\pm1.13$ & $1489.32\pm125.24$ & $1489.32\pm125.24$\\
        LYPPO  & $\mathbf{2234.58}\pm34.92$  & $1086.27\pm2.12$ & $707.08\pm300.61$ & $41.18\pm6.05$\\
        MAPPO  & $1934.72\pm149.66$& $1086.06\pm0.27$ & $1880.80\pm155.49$ & $1549.06\pm271.32$\\
        NCLF & $1887.36\pm26.98$ & $1078.19\pm660.76$ & $1979.02\pm8.03$ & $871.87\pm46.02$\\
        \bottomrule 
    \end{tabular}
    \caption{The expected reward of \algo\ and the baselines in the small-scale environments}
    \label{tab:result-small}
\end{table*}

\begin{figure*}[t]
    \centering
    \subfigure[\small{Microgrid5}]{\includegraphics[width=.16\textwidth]{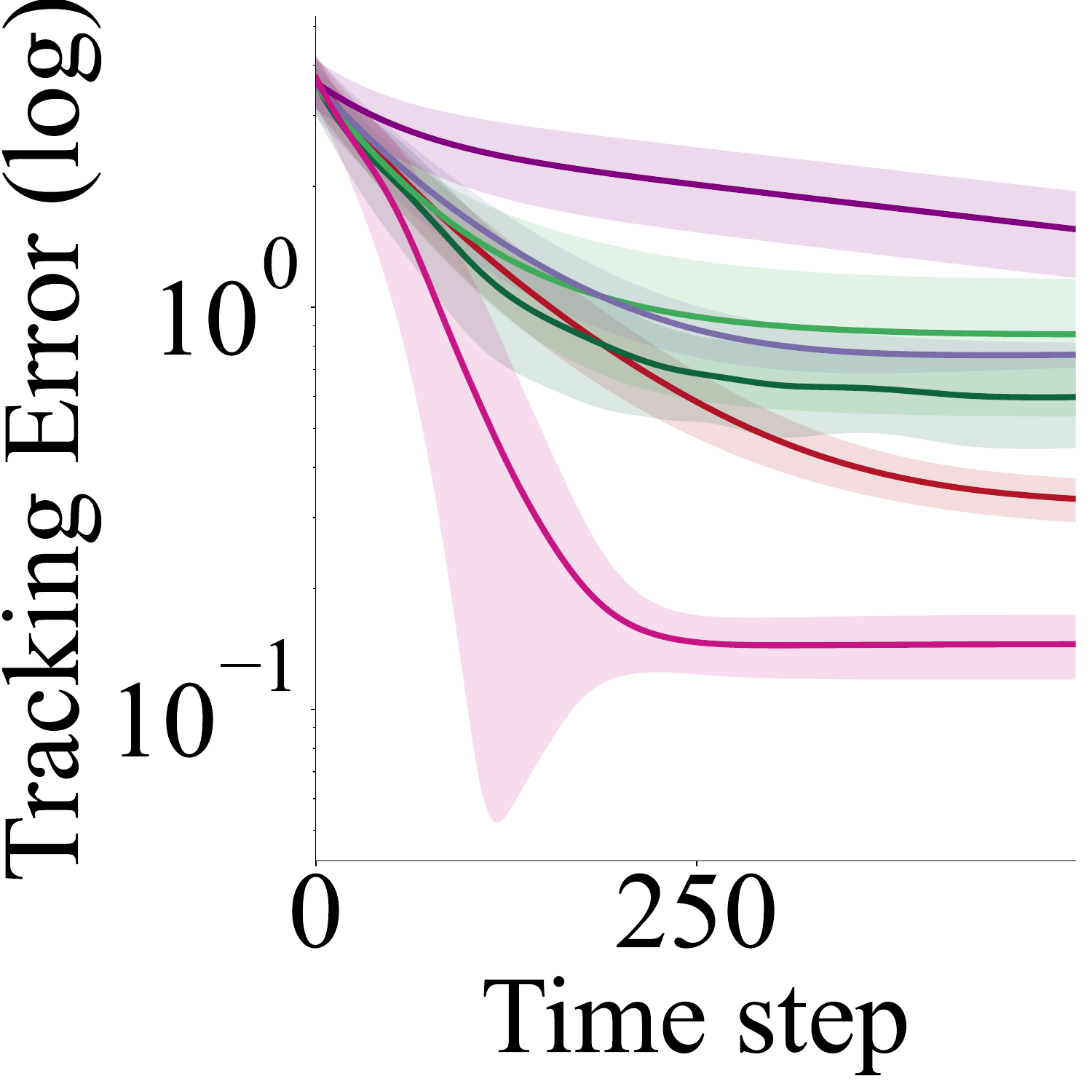}}
        \subfigure[\small{GridVoltage8}]{\includegraphics[width=.16\textwidth]{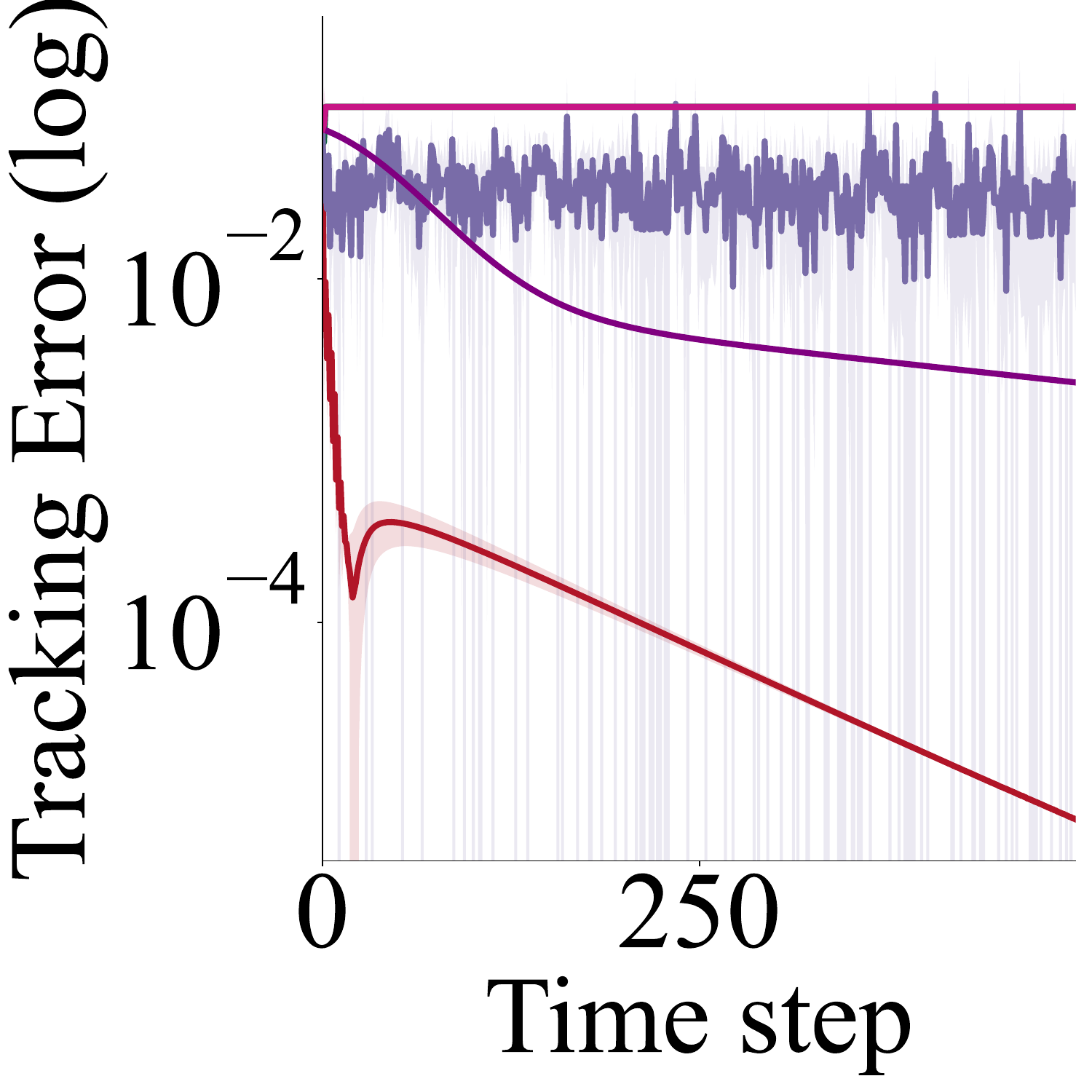}}
    \subfigure[\small{Platoon5}]{\includegraphics[width=.16\textwidth]{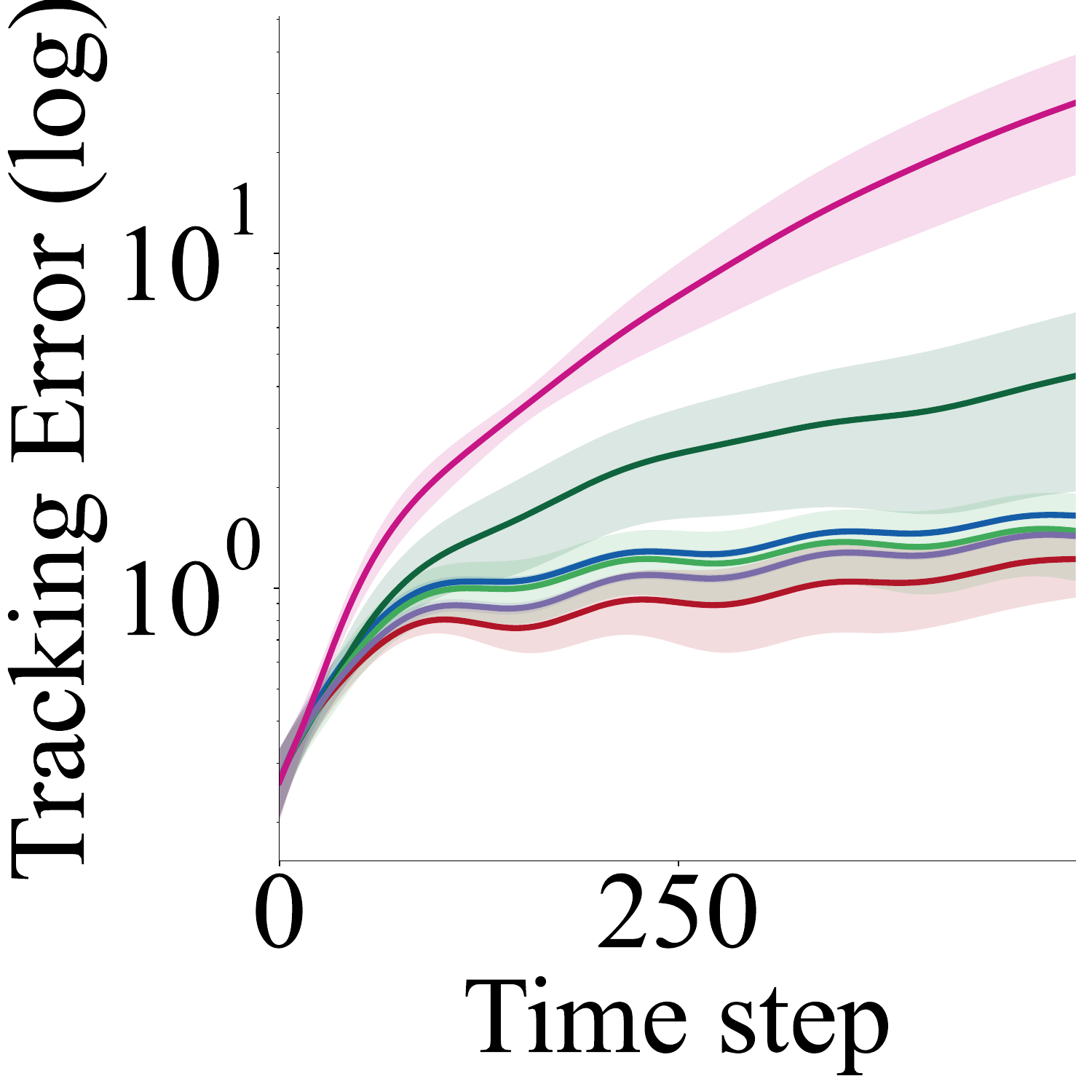}}
    \subfigure[\small{Drone2x2}]{\includegraphics[width=.16\textwidth]{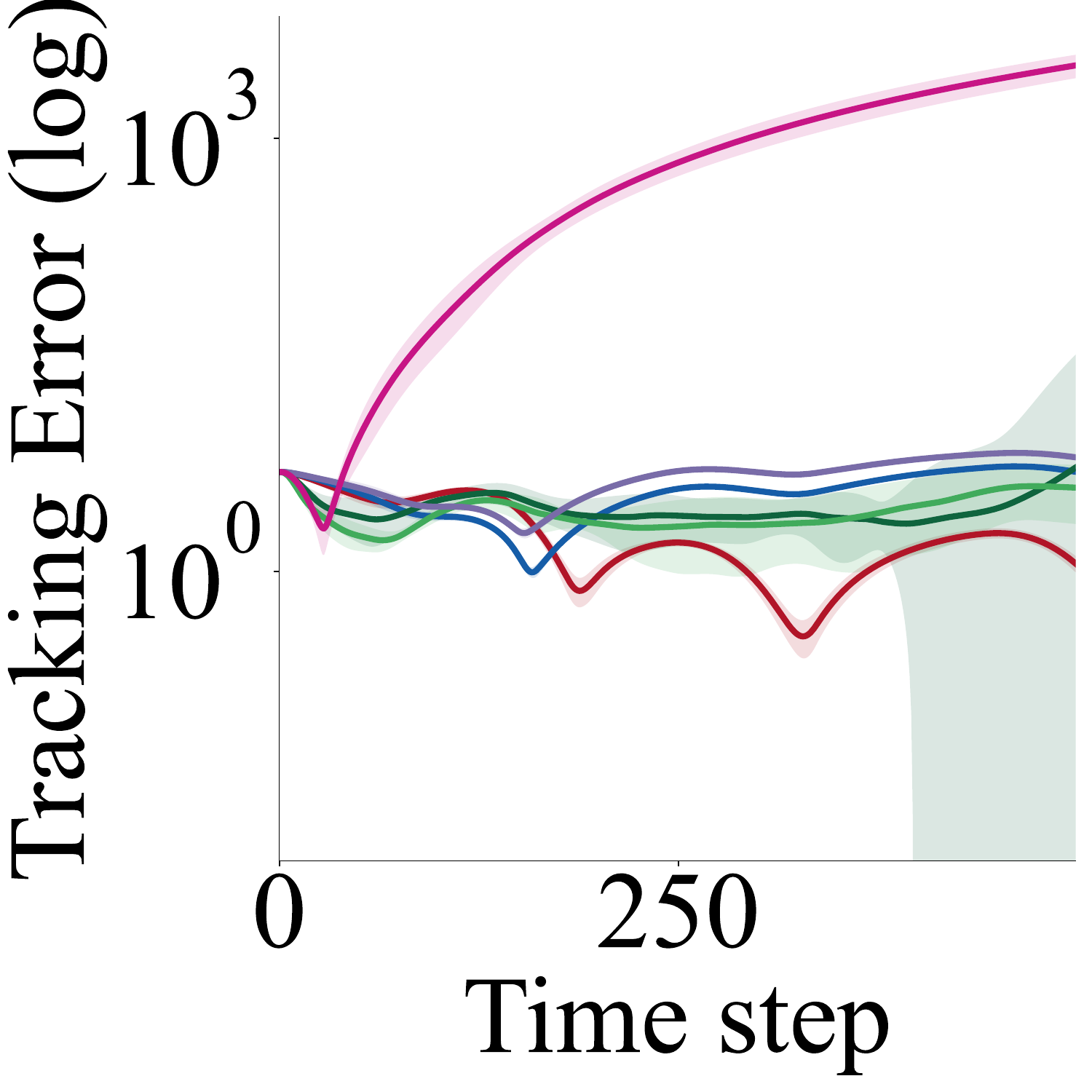}}
    \subfigure[\small{Platoon100}]{\includegraphics[width=.16\textwidth]{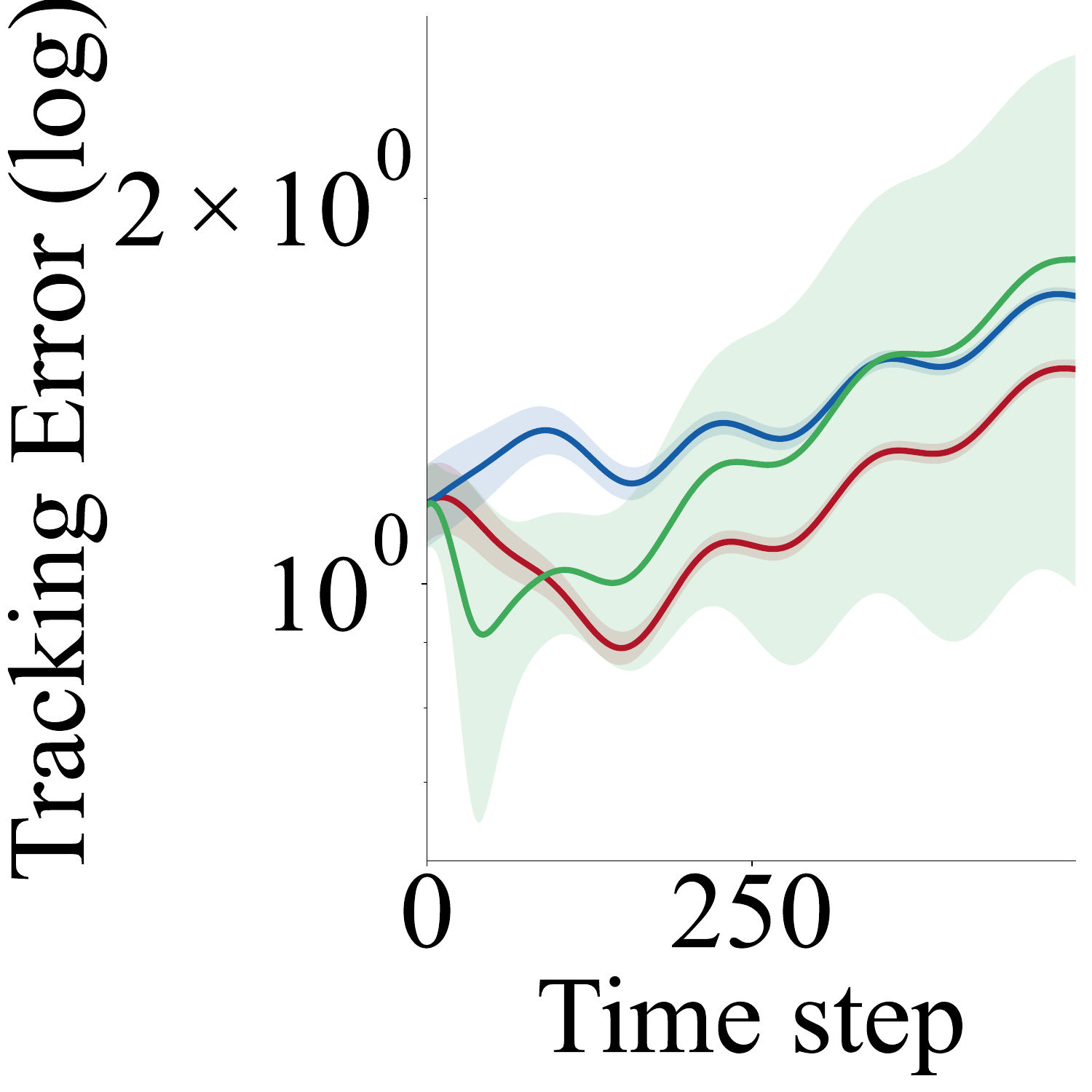}}
    \subfigure[\small{Drone10x10}]{\includegraphics[width=.16\textwidth]{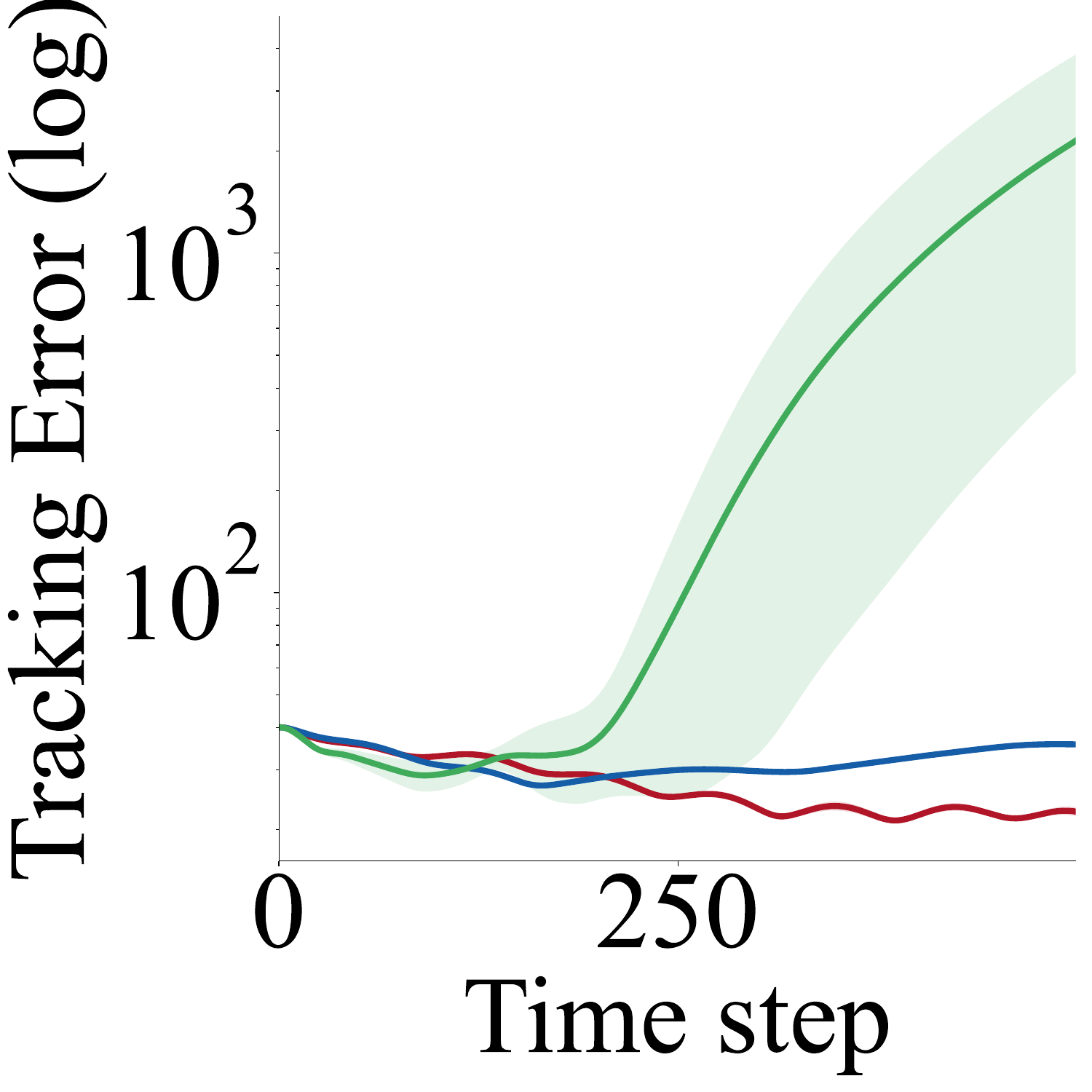}}
    \includegraphics[width=.99\textwidth]{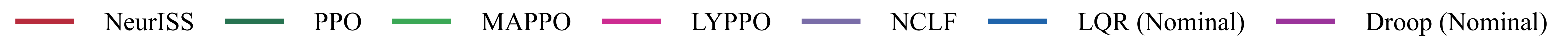}
    \vspace{-8pt}
    \caption{The tracking error in log scale w.r.t. time step of \algo\ and the baselines in the small-scale environments (\textit{a-d}) and the large-scale environments (\textit{e-f}). The line shows the mean tracking error while the shaded region shows the standard deviation.}
    \label{fig:error}
\end{figure*}

\subsection{Results}

The results show that compared with the baselines, \algo\ can achieve comparable or better rewards and tracking errors in small-scale environments, and significantly higher rewards and lower tracking errors in large-scale environments, which demonstrate its efficacy and generalizability.

\textbf{Small-scale Experiments.}
In \Cref{tab:result-small}, we show the expected rewards and standard deviations of \algo\ and the baselines, and in \Cref{fig:error} (a-d) we show the tracking error w.r.t. the simulation time steps. 
{We can observe that in the small-scale system Microgrid5 ($10$ dimensions), \algo\ achieves the second highest expected reward, and second lowest tracking error, while LYPPO behaves the best. 
In GridVoltage8 (8 dimensions) and larger systems Platoon5 and Drone2x2 ($15$ and $24$ dimensions), \algo\ achieves the highest expected rewards and the lowest tracking error. Note that in Platoon5 and Drone2x2, }we do not have full knowledge of the tracking trajectory, and the trajectory changes fast, so the tracking error cannot converge to $0$. \algo\ has this performance because of its ability to learn decentralized controllers \emph{jointly} with the ISS Lyapunov functions as certificates. Compared with NCLF, \algo\ performs better because it is hard to find a global CLF for networked systems. PPO and MAPPO achieve lower rewards than \algo.
% \footnote{{For GridVoltage8, PPO, MAPPO, and LYPPO have almost identically bad rewards and tracking errors. This is because we use a professional solver PandaPower \citep{pandapower} to simulate the underlying distribution grid. All three RL methods return controllers that quickly drive the system to an unsafe state, making the solver fail to solve the underlying model. In such cases, we simply set the reward and the tracking error assuming a 10\% voltage deviation, which is a typical safety limit for real-world distribution grids \citep{shi2022stability}. }} 
They are policy gradient methods to approximate the solution of the Bellman equation, so there is no certificate of their stability. LYPPO, although outperforms \algo\ in very small systems (Microgrid), its performance drops a lot in larger systems (Platoon and Drone). This is because in small-scale systems, with the guidance of CLF, RL can achieve the goal very quickly to maximize the cumulative reward, but \algo\ only seeks to reach the goal without targeting on the convergence speed. Therefore, \algo\ convergences slower than LYPPO. However, in larger scale systems, because of the hardness of finding a correct global CLF, LYPPO receives the wrong guidance by the wrong CLF, and thus behaves much worse than \algo\ and even PPO and MAPPO. 
For the nominal controllers, LQR is designed for linear systems and the Droop controllers are hand-tuned. Therefore, their performance is hard to guarantee in complex nonlinear networked systems. 

\textbf{Large-scale Experiments.}
One key advantage of the proposed framework is network generalizability (\Cref{subsec:generalizability}), where the decentralized controllers and ISS Lyapunov functions trained in small networked systems can be directly applied to large networked systems without further training, while the centralized controllers need a really long time to be trained on large-scale systems (Approximately 250 hours for Drone10x10). We test the $3$ decentralized approaches, \algo, MAPPO, and LQR in large-scale Platoon and Drone systems, Platoon100 and Drone10x10, with $100$ trucks and $10\times10=100$ drones. We show the tracking errors w.r.t. the simulation time steps in \Cref{fig:error} (e-f). We observe that \algo\ has the smallest tracking errors in both environments, with large gaps to others, which shows that \algo\ has the strongest scalability.

%===============================================================================

\section{Conclusion}\label{sec:conclusion}
In this paper, we propose a neural compositional certificate framework for stabilizing large-scale networked dynamical systems. Limitations of the approach include: 1) the approach requires the knowledge of the dynamical system functions $f_i$; 2) the network generalizability result \Cref{lem:generalizaion} requires a strong symmetric condition; 3) the robust ISS Lyapunov result \Cref{lem:robust-iss} assumes the dynamical system is control affine; 4) the approach only learns a compositional certificate using finite samples but does not verify it, so in some sense, the ISS Lyapunov functions we learn are only candidate ISS Lyapunov functions. These limitations are all interesting future directions.

% Acknowledgments---Will not appear in anonymized version
\acks{The Defense Science and Technology Agency in Singapore and the C3.ai Digital Transformation Institute provided funds to assist the authors with their research. Guannan Qu is also supported by NSF Grant 2154171. However, this article solely reflects the opinions and conclusions of its authors and not DSTA Singapore, the Singapore Government, or C3.ai Digital Transformation Institute.}

\bibliography{main.bib}

\newpage

\appendix

\section{Lyapunov Function}\label{sec:Lyapunov}

Lyapunov functions are widely used to guarantee the stability of the \songyuan{dynamical systems}. Lyapunov functions are formally defined in the following \Cref{prop:lyapunov}.
\begin{proposition}
\label{prop:lyapunov}
Given a dynamical system $\dot{x} = \bar{f}(x)$ where $\bar{f}:\mathbb{R}^d \rightarrow \mathbb{R}^d$, suppose there exists a differentiable and radially unbounded function $V:\mathbb{R}^d\rightarrow\mathbb{R}$ that satisfies the following conditions: 
\begin{align*}
     V(x) = 0 \; \forall x\in\xg, \; V(x)>0 \; \forall x\not\in \xg,\text{ and } \nabla V(x)^\top \bar{f}(x) <0 \; \forall x\not\in \xg. 
\end{align*}
Then the system is asymptotically stable \songyuan{about} $\xg$, and $V(x)$ is called a Lyapunov function.
\end{proposition}

A common paradigm for stabilizing a dynamical system is to jointly search for a controller $u=\pi(x)$ and a Lyapunov function $V(x)$ that satisfies the conditions in \Cref{prop:lyapunov}. 
\songyuan{However, finding a Lyapunov function for a large networked system is not trivial when the dimension of the state space (i.e. the input to the Lyapunov function) increases with the number of subsystems.} In this paper, we utilize a decentralized compositional Lyapunov approach to improve the scalability of Lyapunov-based methods for large-scale system control.

\section{Proofs} \label{sec:proofs}
In this section, we provide proof for the lemmas stated in the main content of the paper. 

\subsection{Proof of \Cref{lem:iss_lyapunov}}

\begin{proof}
The proof follows similar steps as that in \citet{liu2011lyapunov} and we modify it to accommodate the network structure in our setting. Recall that we use $f(x,u)$ to denote the vector of the individual dynamical functions $f_i$. Further, we use $\pi(x)$ to denote $\pi(x) = [\pi_1(x_1),\ldots,\pi_n(x_n)]^\top $. With this notation, the closed-loop dynamical system can be written as 
\[ \dot{x} = f(x,\pi(x)). \]

Now consider the following function 
\[V(x) = \max_{i\in\mathcal{N}} V_i(x_i). \]

\textbf{Case 1. } Suppose for a given $x$, $\max_{i\in\mathcal{N}} V_i(x_i)$ is uniquely achieved at $i^*\in\mathcal{N}$. Due to the continuous differentiability of the $V_{i}$'s, we have $V(x') = V_{i^*} (x'_{i^*})$ for $x'$ in a neighborhood of $x$, and as a result, $V(x)$ is continuously differentiable at $x$. 
Further, note that clearly $V_{i^*}(x_{i^*}) > \max_{j\in\mathcal{N}_{i^*}} V_j(x_j) \geq \max_{j\in\mathcal{N}_{i^*}} \chi_{{i^*}j}(V_j(x_j) )$. Using the imply condition \eqref{eq:imply}, we have,
\begin{align}
   & [\nabla V(x)]^\top f(x,\pi(x))\nonumber\\
    &=[\nabla V_{i^*}(x_{i^*}) ]^\top f_{i^*}(x_{i^*},x_{\mathcal{N}_{i^*}},\pi_{i^*}(x_{i^*}))\nonumber \\
    &\leq -\alpha_{i^*} V_{i^*}(x_{i^*}) = -\alpha_{i^*} V(x).  \label{eq:proof_iss_case1}
\end{align}

\textbf{Case 2. } Suppose for a given $x$, $\max_{i\in\mathcal{N}} V_i(x_i)$ is achieved at a set of multiple indices $\mathcal{I}$. Then, following a similar argument as Case 1 and using a continuity argument, there must exist a neighborhood $\mathcal{A}$ around $x$ and constant $\lambda>0$ such that for all $j\in \mathcal{I}$ and for all $y\in \mathcal{A}$,
\begin{align}
    [\nabla V_{j}(y_{j}) ]^\top f_{j}(y_{j},y_{\mathcal{N}_{j}},\pi_{j}(y_{j})) 
    &\leq -\lambda V_{j}(x_{j}), \label{eq:proof_iss_case2_neighborhood}
\end{align}
and further, $\max_{i\in\mathcal{N}} V_i(y_i)$ must be achieved within $\mathcal{I}$. 

Consider a trajectory of the system $\dot{y} = f(y,\pi(y))$ starting at $y(0)=x$. There must exist $\delta>0$ s.t. $y(t)\in\mathcal{A},\forall t\in[0,\delta]$. For any $t$ within range $[0,\delta]$,  suppose $\max_{i\in\mathcal{N}} V_i(y_i(t))$ is achieved at a certain $j\in\mathcal{I}$. Then, we have,
\begin{align*}
&    V(y(t)) - V(x) \\
    & = V_j (y_j(t)) - V_j(y_j(0))\\
    &=\int_{\tau=0}^t \frac{d}{d\tau}  V_j(y_j(\tau)) d\tau \\
    &= \int_{\tau=0}^t [\nabla  V_j(y_j(\tau))]^\top f_j(y_j(\tau),y_{\mathcal{N}_j}(\tau),\pi_j(y_j(\tau))) d\tau\\
    &\leq -\lambda \int_{\tau=0}^t V_j(x_j) d\tau \\
    &= -\lambda V_j(x_j) t = -\lambda V(x) t\end{align*} 
    where in the last inequality we have used \eqref{eq:proof_iss_case2_neighborhood}.
Therefore, whenever $V$ is differentiable at $x$, we can divide the above by $t$ and let $t\rightarrow 0$ to get
\begin{align}
    [\nabla V(x)]^\top f(x,\pi(x))
    \leq -\lambda V(x).
\end{align}
Combining the above and \eqref{eq:proof_iss_case1}, we have there exists a constant $\lambda'>0$ s.t. $[\nabla V(x)]^\top f(x,\pi(x))
    \leq -\lambda' V(x)$ whenever $V$ is differentiable. Also, it is easy to check that $V$ is continuously differentiable almost everywhere, and satisfies
    \[ \underline{\alpha}( \dist( x,\xg))\leq V(x)\leq \bar{\alpha}(\dist( x , \xg)),\]
    for $\mathcal{K}_\infty$ functions $\underline{\alpha}$ and $\bar{\alpha}$. 
    Lastly, we can use the same argument as in the proof of \citet[Thm 3.1]{jiang1996lyapunov} to show that the trajectory of the dynamical system must converge to the goal set. 
\end{proof}

\subsection{Proof of \Cref{lem:generalizaion}}

\begin{proof}
Fix a $j\in\tilde{\mathcal{N}}$. For $j'\in \tilde{\mathcal{N}}_j$, we have 
\begin{align}
    \tilde{V}_{j'} = V_{\tau_{j'}(j')} = V_{\tau_j(j')}. \label{eq:scale_up_v}
\end{align}

Define $\tilde{\chi}_{jj'} = \chi_{\tau_j(j)\tau_j(j')}$ and $\tilde{\alpha}_j = \alpha_{\tau_j(j)}$. Suppose $x_j$ and $x_{\tilde{\mathcal{N}}_j}$ are such that 
\begin{align}
    \tilde{V}_j (x_j) \geq \max_{j'\in \tilde{\mathcal{N}}_j} \tilde{\chi}_{j j'} (\tilde{V}_{j'}(x_{j'}) )\label{eq:scale_up_imply1}
\end{align}
Using \eqref{eq:scale_up_v}, we have \eqref{eq:scale_up_imply1} is equivalent to 
\begin{align*}
    V_{\tau_j(j)}(x_j) \geq \max_{ j'\in \tilde{\mathcal{N}}_j} \chi_{\tau_j(j)\tau_j(j')} ( V_{\tau_j(j')} (x_{j'})).
\end{align*}
By \eqref{eq:imply}, we have the above implies,
\begin{equation}
    \begin{aligned}
    [\nabla V_{\tau_j(j)}(x_j)]^\top f_{\tau_j(j)} (x_{j}, x_{\tilde{N}_j}, &\pi_{\tau_j(j)}(x_j) )\\
    &\leq -\alpha_{\tau_j(j)} V_{\tau_j(j)}(x_j),\nonumber
    \end{aligned}
\end{equation}
which is equivalent to 
\begin{equation}\label{eq:scale_up_imply2}
    \begin{aligned}
    [\nabla \tilde{V}_j(x_j)]^\top \tilde{f}_j (x_{j}, x_{\tilde{N}_j}, \tilde{\pi}_j(x_j) )
    \leq -\tilde{\alpha}_j \tilde{V}_j(x_j). 
\end{aligned}
\end{equation}
This shows that \eqref{eq:scale_up_imply1} can imply \eqref{eq:scale_up_imply2}. As such, $\tilde{V}_j$ is a valid compositional certificate for the new system. 
\end{proof}

\subsection{Proof of \Cref{lem:robust-iss}}

\begin{proof}
Using the control-affine dynamics, we have 
\begin{equation}\label{eq:decent-affine}
    \begin{aligned}
    [\nabla &V_i (x_i) ]^\top f_i(x_i,x_{\mathcal{N}_i}, \pi_i(x_i))\\
    &=[\nabla V_i (x_i) ]^\top \left[h_i (x_i,x_{\mathcal{N}_i};\beta) + g_i (x_i,x_{\mathcal{N}_i};\beta)\pi(x_i)\right]\\
    &=L_{h_i(\beta)}V_i(x_i)+L_{g_i(\beta)}V_i(x_i)\pi(x_i),
    \end{aligned}
\end{equation}
where we denote 
\begin{equation}
    \begin{aligned}
        L_{h_i(\beta)}V_i(x_i)&=[\nabla V_i (x_i) ]^\top h_i(x_i,x_{\mathcal{N}_i};\beta)\\
        L_{g_i(\beta)}V_i(x_i)&=[\nabla V_i (x_i) ]^\top g_i(x_i,x_{\mathcal{N}_i};\beta)
    \end{aligned}
\end{equation}
as the Lie derivatives of $V_i$ along $h_i(x_i,x_{\mathcal{N}_i};\beta)$ and $g_i(x_i,x_{\mathcal{N}_i};\beta)$. By assumption, we have $h_i$ and $g_i$ are affine in $\beta$. In addition, the Lie derivatives $L_{h_i(\beta)}V_i$ and $L_{g_i(\beta)}V_i$ are affine in $h_i$ and $g_i$, and \Cref{eq:decent-affine} is affine in $L_{h_i(\beta)}V_i$ and $L_{g_i(\beta)}V_i$. Therefore, the mapping from $\mathcal{B}$ to \Cref{eq:decent-affine} is affine which maps the convex hull of $\beta_1,\beta_2,\ldots,\beta_{n_\beta}$ to the convex hull of $L_{h_i(\beta_1)}V_i+L_{g_i(\beta_1)}V_i\pi_i,\ldots,L_{h_i(\beta_{n_\beta})}V_i+L_{g_i(\beta_{n_\beta})}V_i\pi_i$. As a result, if the conditions in \Cref{lem:iss_lyapunov} are satisfied for $\beta_i,i=1,2,\ldots,n_\beta$, then the conditions are satisfied for any $\beta\in\mathcal{B}$. Using \Cref{lem:iss_lyapunov}, we can conclude that the closed-loop system is globally asymptotically stable with any $\beta\in\mathcal{B}$. 
\end{proof}

\section{Experiment Details}

Here we provide additional experimental details and results. \footnote{We provide the code of our experiments at \href{https://github.com/MIT-REALM/neuriss}{https://github.com/MIT-REALM/neuriss}.} Our experiments are run on a 64-core AMD 3990X CPU @ 3.60GHz and four NVIDIA RTX A4000 GPUs (one GPU each training job). 

\subsection{Environment Details}\label{subsec:environment_detail}

\subsubsection{Power Systems}

\begin{figure}[ht]
    \centering
    \includegraphics[width=.45\textwidth]{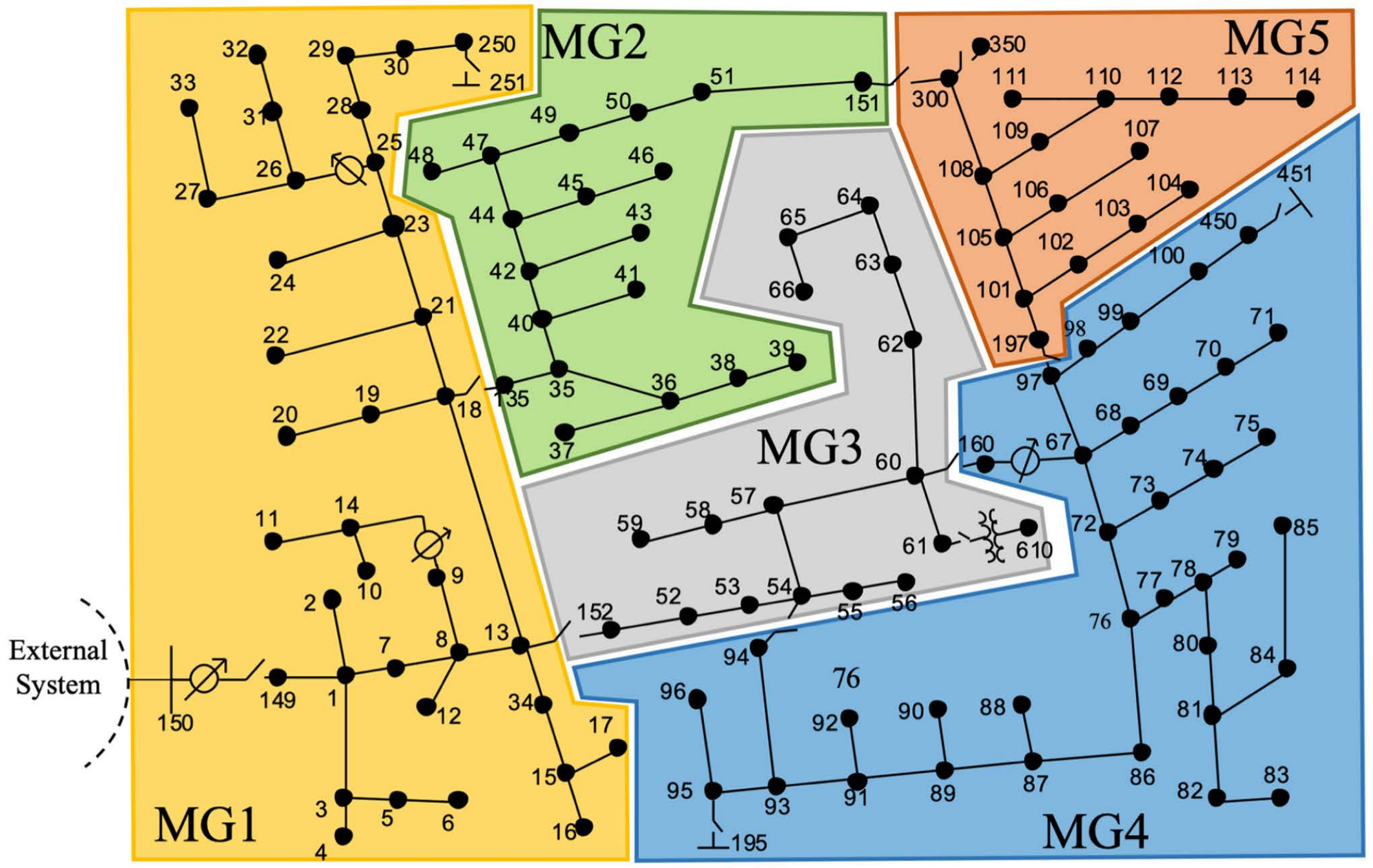}
    \caption{IEEE 123-node Test Feeder~\cite{huang2021neural}}
    \label{fig:ieee-123-illustrate}
\end{figure}

\textbf{Networked Microgrid. }We consider the networked microgrid system introduced by \citet{huang2021neural}, which is shown in \Cref{fig:ieee-123-illustrate}. In this environment, a power distribution network is divided into 5 regions (MG$1$, MG$2$, ..., MG$5$ in \Cref{fig:ieee-123-illustrate}). Each region $i$ functions as a microgrid and two microgrids are neighbors when their corresponding regions are connected by a power line. Each microgrid $i$ has two states $x_i = (\delta_i,E_i)$ where $\delta_i$ means the voltage phase angle and $E_i$ is the voltage magnitude. The dynamics of microgrid $i$ is given by
\begin{subequations}
\begin{align}
    &M_{a,i} \dot{\delta}_i + (\delta_i - \delta_i^{\mathrm{ref}}) = u_i^{P} \nonumber \\
    &\ + D_{a,i} (P_i^{\mathrm{ref}} -G_{ii}E_i^2 -  \sum_{j\in \mathcal{N}_i} E_i E_j Y_{ji}\cos(\delta_j - \delta_i - \sigma_{ji})  ),\\
    & M_{v,i} \dot{E}_i + (E_i - E_i^{\mathrm{ref}}) = u_i^{Q} \nonumber \\
    &\ + D_{v,i} (Q_i^{\mathrm{ref}} +B_{ii}E_i^2 -  \sum_{j\in \mathcal{N}_i} E_i E_j Y_{ji}\sin(\delta_j - \delta_i - \sigma_{ji})  ),
\end{align}
\end{subequations}
where $M_{a,i}$ and $M_{v,i}$ are innertia coefficients, $D_{a,i}$ and $D_{v,i}$ are droop coefficients to provide internal droop controls, $\delta_i^{\mathrm{ref}} ,E_i^{\mathrm{ref}}, P_i^{\mathrm{ref}}, Q_i^{\mathrm{ref}}$ are precomputed reference values, $B_{ii}, G_{ii}$, $Y_{ji}$, $\sigma_{ji}$ are coefficients from the network admittance matrix. For a detailed description of the model, see \citet{huang2021neural}. Note that compared to \citet{huang2021neural}, we also introduce the control input $u_i = (u_i^P, u_i^Q)$ which represents the active and reactive power produced by the secondary control, and the goal is to design secondary controllers so that $\delta_i, E_i$ can converge to their reference values $\delta_i^\mathrm{ref},E_i^\mathrm{ref}$. 

During training, the training data are sampled from $\delta_i\in[-3,3]$ and $E_i\in[-3,3]$, and the data in the goal region are sampled from the region where $\delta_i=\delta_i^\mathrm{ref}$ and $E_i=E_i^\mathrm{ref}$. During testing, we set the simulation time interval $\Delta t=0.01$, and randomly sample the initial states of the microgrids in $\delta_i\in[-2,2]$ and $E_i\in[-3,3]$. The number of simulation time steps is $500$.

\textbf{Distribution grid voltage control. } We consider a power system voltage control problem given in \cite{shi2022stability}. There is a power distribution network as a graph $\mathcal{G} = (\mathcal{N},\mathcal{E})$, which consists of a set of nodes $\mathcal{N}=\{1,\ldots,n\}$ and edges $\mathcal{E}$. 
Each node $i\in\mathcal{N}$ is associated with a reactive power injection $q_i$, and a voltage magnitude $v_i$. We use $q$ and $v$ to denote the $q_i,v_i$ stacked into a vector. The system dynamics is given as follows
\begin{align}
\dot{q}(t) = u(t) = \pi(v(t))\label{eq_rl:policy}\\
v(t) = v(q(t))
\end{align}
where the state is the reactive power $q(t)$, and the control action is the change rate of $q(t)$.
Critically, the voltage $v(t)$ is a function of the reactive power $q(t)$, denoted as $v(t) = v(q(t))$, and this function is defined implicitly via the solution of a nonlinear algebraic equation system known as the power-flow equation \citep{low2014convex}. In our experiment, we use PandaPower \citep{pandapower} as the powerflow solver. The goal is to design a controller $u(t) = \pi(v(t))$ where the control action $u(t)$ depends on the voltage $v(t)$ such that in the close loop system, the voltage $v(t)$ across all nodes will converge to the nominal value (which is $1.0$). In other words, the goal points of the problem consist of the set of reactive power such that the voltage is $1.0$. 

\begin{figure}[ht]
    \centering
    \includegraphics[width=.45\textwidth]{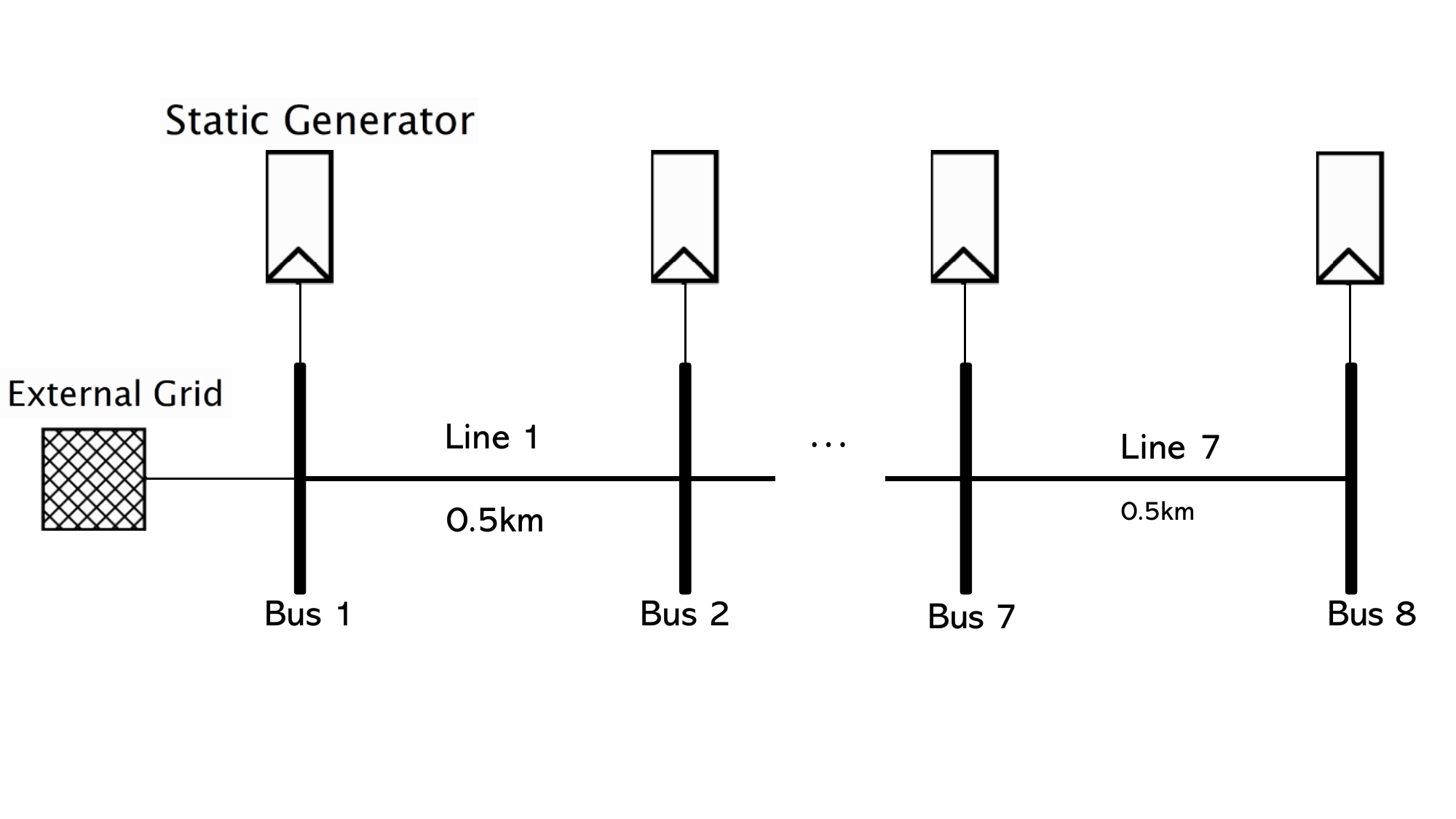}
    \caption{GridVoltage8}
    \label{fig:8bus}
\end{figure}

In our experiments, we consider a power distribution system that consists of $8$ buses, see \Cref{fig:8bus}. The buses are arranged in a line, each one is connected to a static generator. The nominal controller is similar to the droop control and is a proportion controller on the voltage deviation, $u_i(t) = c_i(v_i(t)-1)$ (where $c_i$ is a constant). This proportion controller is a standard controller used in practice.  

During training, the training data are sampled from $q_i\in[-0.001,0.001]$, since Pandapower will give no solution when the variance of the training data is too large, in this case, we manually set the output voltage $v_i(t)$ to 0.9 for negative states $q_i(t)$ and 1.1 for positive ones. During testing, we set the simulation time interval $\Delta t=0.01$, and sample the initial states of the power system from states of all 0. The number of simulation time steps is 500.

\subsubsection{Platoon}

\begin{figure}[ht]
    \centering
    \includegraphics[width=.45\textwidth]{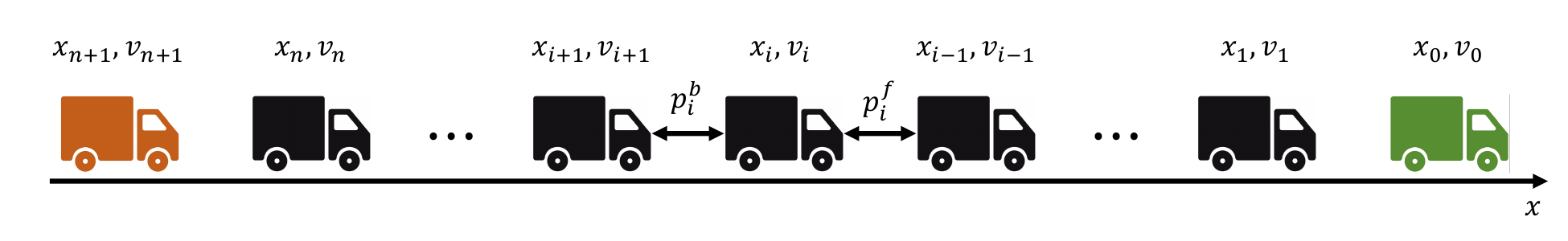}
    \caption{Platoon}
    \label{fig:platoon-illustrate}
\end{figure}

A platoon system is shown in \Cref{fig:platoon-illustrate}, which contains a list of trucks. The green truck is the leading truck (the $0$-th truck), which can drive freely, and the orange truck is the last truck (the $(n+1)$-th truck). We want to control the trucks in the middle (the black trucks) so that the trucks in the whole truck platoon system are spread evenly. As mentioned in the main pages, the state of the $i$-th truck is defined as $x_i=[p_i^f,p_i^b,v_i]^\top$, where $p_i^f$ is the distance between the $i$-th truck and the $(i-1)$-th truck, $p_i^b$ is the distance between the $i$-th truck and the $(i+1)$-th truck, and $v_i$ is the velocity of the $i$-th truck. The state of the $i$-th truck is shown in \Cref{fig:platoon-illustrate}. For each relative position of the leading and last trucks, the goal set for each truck $i\in\{1,2,...,n\}$ is uniquely defined as the set of states satisfying $p_i^f=p_i^b$, which means that each middle truck aims to keep the distance to the front and behind trucks the same.

\begin{figure}[ht]
    \centering
    \includegraphics[width=.45\textwidth]{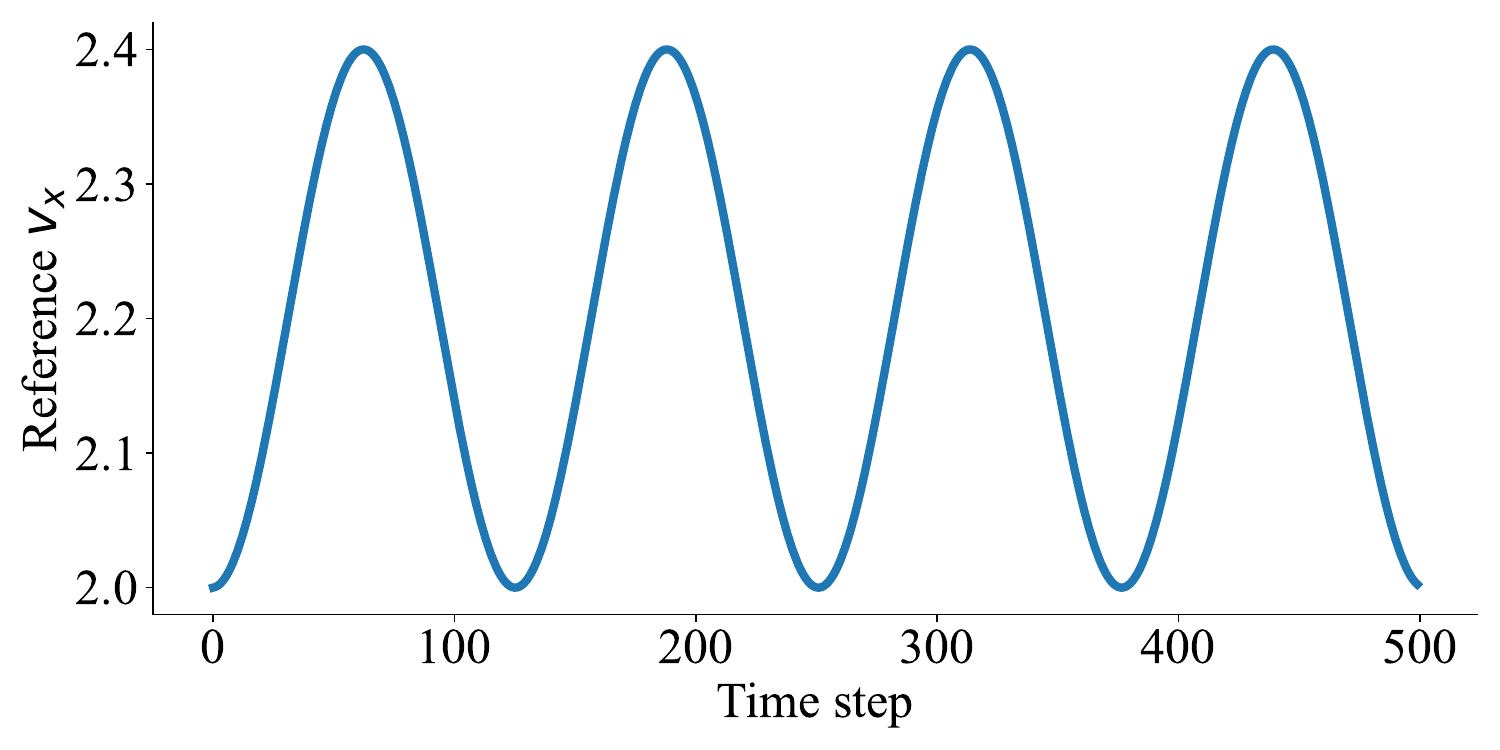}
    \caption{The velocity profile of the leading truck of the platoon system}
    \label{fig:platoon-vx-ref}
\end{figure}

During training, the training data are sampled from $p_i^f\in[0,2]$, $p_i^b\in[0,2]$, $v_i\in[0,4]$, and the data in the goal region are sampled from the region where $p_i^f=p_i^b$. During testing, the leading truck follows a trajectory with initial velocity $2.0$ and acceleration $\sin(5t\Delta t)$, where $t$ is the simulation time step and $\Delta t=0.01$ is the simulation time interval. Therefore, the velocity of the leading truck follows the profile shown in \Cref{fig:platoon-vx-ref}. We randomly sample the controlled trucks' initial states in $p_i^f\in[0.6,1.4]$, $p_i^b\in[0.6,1.4]$, and $v_i\in[1.0,1.2]$, so that the trucks need to speed up first to catch the leading truck. The number of simulation time steps is $500$.

\subsubsection{Planar Drone Formation Control}

\begin{figure}[ht]
    \centering
    \includegraphics[width=.3\textwidth]{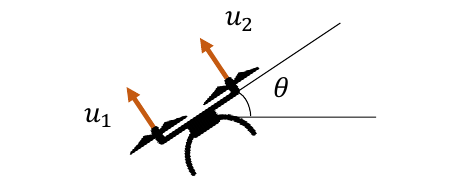}
    \caption{Planar Drone Dynamics \citep{underactuated}}
    \label{fig:drone-dyn}
\end{figure}

A Planar Drone is shown in \Cref{fig:drone-dyn} \citep{underactuated}. For a single drone, the state is given by $x=[p_x,p_y,\theta,v_x,v_y,\omega]\top$, where $(p_x,p_y)$ is the position, $(v_x,v_y)$ is the velocity, $\theta$ is shown in \Cref{fig:drone-dyn}, and $\omega$ is the changing rate of $\theta$. The control inputs are forces generated by the two propellers $u_1,u_2$ shown in \Cref{fig:drone-dyn}. The dynamics of the drone is given by $\dot x=h(x)+g(x)u$, where
\begin{equation}
    h(x)=\left[\begin{array}{c}
        v_x \\ v_y \\ \omega \\ 0 \\ -g \\ 0
    \end{array}\right],
\end{equation}
and
\begin{equation}
    g(x)=\left[\begin{array}{cc}
        0 & 0 \\ 0 & 0 \\ 0 & 0 \\ 
        -\frac{\sin\theta}{m} & -\frac{\sin\theta}{m} \\ 
        \frac{\cos\theta}{m} & \frac{\cos\theta}{m} \\ 
        \frac{r}{I} & -\frac{r}{I}
    \end{array}\right],
\end{equation}
where $m,I,r$ is the mass, moment of inertia, and the distance from the center to the base of the propeller, respectively.

For the formation control task, we design the states of the drones to be a 2-D platoon-like system, given by $x_i=[p_i^l,p_i^r,p_i^u,p_i^d,\theta_i,v_i^x,v_i^y,\omega_i]^\top$, where $p_i^l,p_i^r,p_i^u,p_i^d$ are the distances from drone $i$ to the left, right, up, down drones. Other dimensions of the state are not coupled and are the same as the single drone system. The control inputs of each drone are the same as the single drone system. 

\begin{figure}[ht]
    \centering
    \includegraphics[width=.45\textwidth]{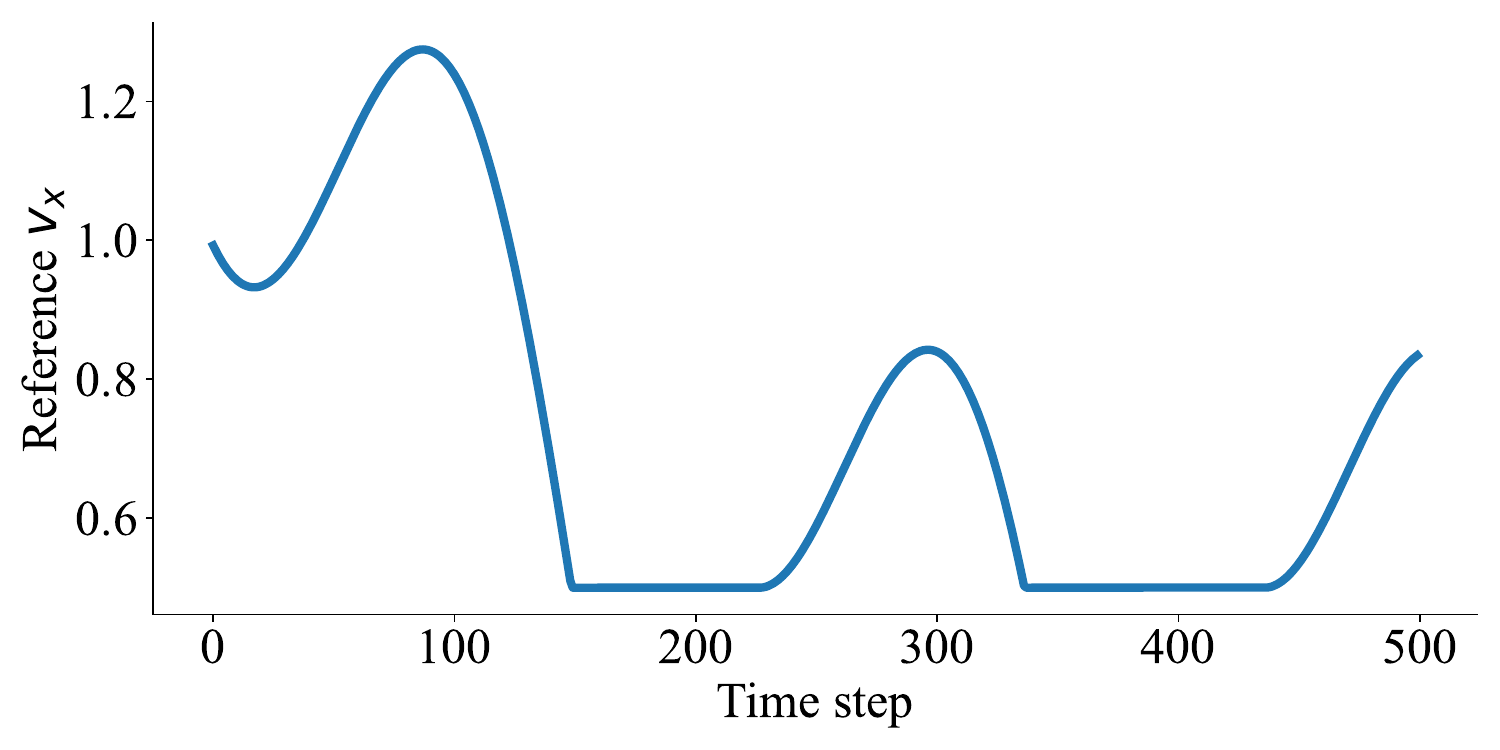}
    \caption{The velocity profile of the tracking trajectory of the planar drone formation control system}
    \label{fig:drone-vx-ref}
\end{figure}

During training, the training data are sampled from $p_i^l,p_i^r,p_i^u,p_i^d\in[0,5]$, $\theta_i\in[-\pi/2,\pi/2]$, $v_i^x\in[-7,7]$, $v_i^y\in[-5,5]$, $\omega_i\in[-\pi/2,\pi/2]$, and the data in the goal region are sampled from the region where $p_i^l=p_i^r,p_i^u=p_i^d$, and $\theta_i=\omega_i=0$. During testing, the trajectory that we want the drones to track follows the profile with initial velocity $1.0$, and acceleration $0.5\sin(t\Delta t)-0.25$, and the velocity is clipped below by $0.5$. Here $t$ is the simulation time step, and $\Delta t=0.03$ is the simulation time interval. Therefore, the velocity of the tracking trajectory follows the profile shown in \Cref{fig:drone-vx-ref}. We randomly sample the drones' initial states in $p_i^l,p_i^r\in[0.8,1.2]$, $p_i^u,p_i^d\in[0.09,0.11]$, $v_i^x\in[0.85,1.15]$, $v_i^y\in[-0.15,0.15]$, $\theta_i\in[-0.05,0.05]$, $\omega_i\in[-0.05,0.05]$, so the drones need to first rise up and then follow the desired trajectory. The number of simulation time steps is $500$.

\subsection{Implementation Details and Additional Results}

\begin{table*}[t]
    \centering
    \begin{tabular}{c|ccccccc}
        \toprule
        Hyper-parameter & $\alpha_i$ & $\epsilon_A$ & $\epsilon_B$ & $\mu_\mathrm{goal}$ & $\mu_{A_i}$ & $\mu_{B_i}$ & $\mu_\mathrm{ctrl}$ \\
        \midrule
        Microgrid & 0.5 & 1.0 & 1.0 & 10 & 0.1 & 50.0 & 0.0\\
        GridVoltage8 & 0.5 & 1.0 & 1.0 & 100 & 0.01 & 50.0 & 1.0\\
        Platoon & 1.0 & 1.0 & 1.0 & 100  & 0.1 & 50 & 0.001 \\
        PlanarDrone & 0.2 & 1.0 & 1.0 & 100 & 0.01 & 3.0 & 0.2\\
        \bottomrule
    \end{tabular}
    \caption{Hyper-parameters used during training \algo}
    \label{tab:hyperparam}
\end{table*}

\begin{table*}[t]
    \centering
    \begin{tabular}{c|cc}
    \toprule
        Environment & Platoon100 & PlanarDrone10x10 \\
        \midrule
        \algo & $\mathbf{1920.20}\pm32.87$ & $\mathbf{44723.37}\pm39.58$ \\
        LQR & $1812.63\pm2.80$ & $44184.47\pm7.38$ \\
        MAPPO & $1816.53\pm177.55$ & $26594.20\pm11563.69$ \\
        \bottomrule
    \end{tabular}
    \caption{The expected reward of \algo\ and the baselines in the large scale environments}
    \label{tab:reward-large}
\end{table*}

In this section, we provide implementation details of \algo\ and the baselines, including the network structures, frameworks or the packages used, and the choice of the optimizer. We also provide the training details including the choice of the hyper-parameters, batch size, number of iterations, random seeds, and other mechanisms used in training. Moreover, we provide additional numerical results of the rewards of \algo\ and the baselines in large-scale environments, and the contour plots of the learned ISS Lyapunov functions. 

\subsubsection{Implementation of \algo}

In our framework, there are two models to be trained: the neural ISS Lyapunov functions $V_i(x_i;S_i,\omega_i,\nu_i)$ and the controllers $\pi_i(x_i;\theta_i)$, where
\begin{equation}
    \begin{aligned}
    V_i(x_i;S_i,\omega_i,\nu_i) &= x_i^\top S_i^\top S_i x_i \\
    &+ p_{i}(x_i;\omega_i)^\top p_{i}(x_i; \omega_i) + q_{i}(x_i;\nu_i),
    \end{aligned}
\end{equation}
and $\pi_i(x_i;\theta_i)$ is a fully-connected multi-layer perceptron (MLP). $S_i\in\mathbb{R}^{d_i\times d_i}$ is a matrix of trainable parameters. The neural networks $p_i(x_i;\omega_i)$ and $\pi_i(x_i;\theta_i)$ are MLPs with two hidden layers with size $64$ and $\mathrm{Tanh}$ as the hidden activation function. $q_i(x_i;\nu_i)$ is an MLP with two hidden layers with size $64$, $\mathrm{Tanh}$ as the hidden activation function, and $\mathrm{ReLU}$ as the output activation function. To control the Lipschitz of the neural networks, we add the spectral normalization mechanism \citep{miyato2018spectral} on every hidden layer of the neural networks. Our framework is implemented in PyTorch~\citep{paszke2019pytorch} framework with ADAM~\citep{kingma2014adam} as the optimizer. 

\subsubsection{Implementation of the Baselines}

Our baselines include the droop controller, LQR \citep{kwakernaak1974linear}, PPO \citep{schulman2017proximal}, LYPPO \citep{chang2021stabilizing}, MAPPO \citep{yu2021surprising}, and neural CLF \citep{dawson2021charles}. The droop controller is only used in \songyuan{Microgrid5 and GridVoltage8}. There is an internal droop controller given with the IEEE 123-node test feeder \citep{huang2021neural} so we directly use that controller as one of the baselines. The droop controller used in\songyuan{ GridVoltage8} is similar to a proportion controller on the voltage deviation, $u_i(t) = c_i(v_i(t)-1)$ (where $c_i$ is a constant), this proportion controller is a standard controller used in practice. The LQR controller is used in the truck platoon system and the planar drone formation control system because these two systems are separable. We linearize the dynamics of the single truck and the single planar drone to calculate the LQR controller for them. PPO is implemented based on the open-source python package stablebaselines\footnote{https://github.com/DLR-RM/stable-baselines3} \citep{stable-baselines3}. LYPPO is implemented based on the official implementation. MAPPO is implemented based on the official implementation\footnote{https://github.com/marlbenchmark/on-policy} \citep{yu2021surprising}. Neural CLF is implemented based on the official implementation\footnote{https://github.com/MIT-REALM/neural\_clbf} \citep{dawson2021charles}. 

\begin{figure*}
    \centering
    \subfigure[Microgrid5]{\includegraphics[width=.45\textwidth]{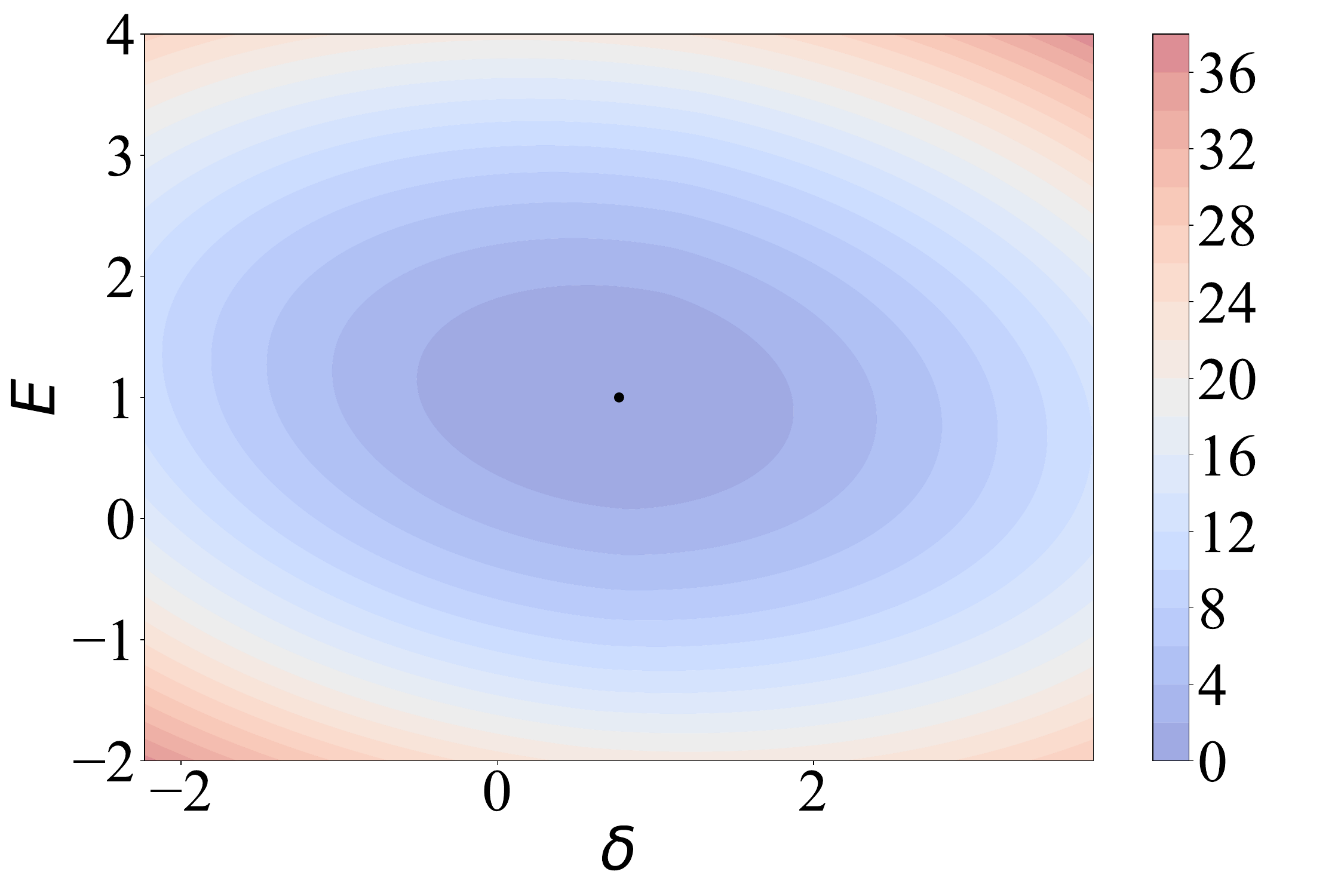}\label{fig:iss-contour-mg5}}
    \subfigure[GridVoltage8,$i=1,...,8$]{\includegraphics[width=.45\textwidth]{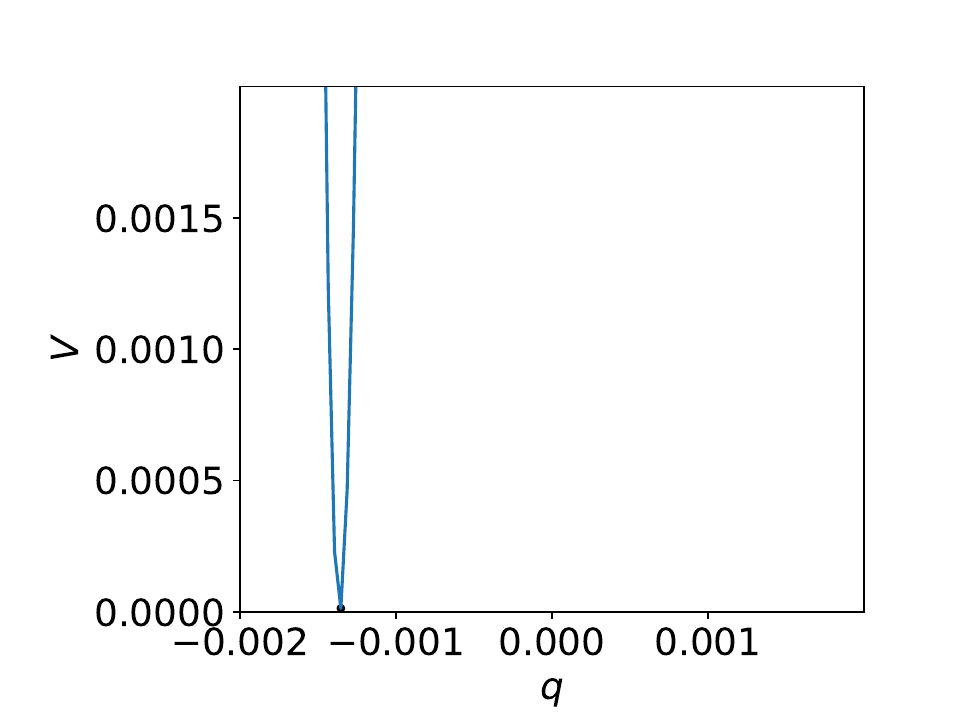}\label{fig:iss-contour-gv8}}
    \subfigure[Platoon5, $i=1,5$]{\includegraphics[width=.45\textwidth]{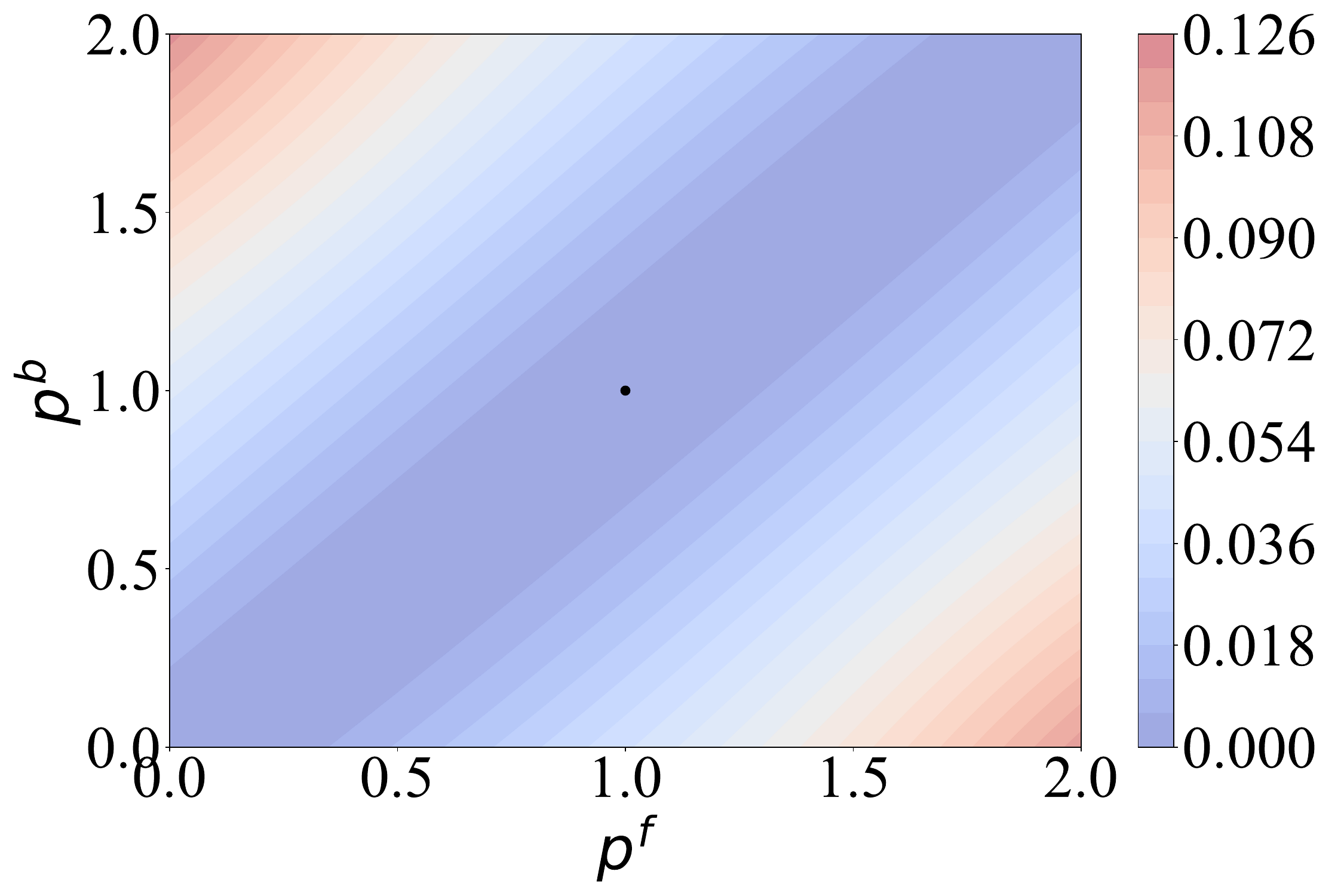}\label{fig:iss-contour-platoon0}}
    \subfigure[Platoon5, $i=2,3,4$]{\includegraphics[width=.45\textwidth]{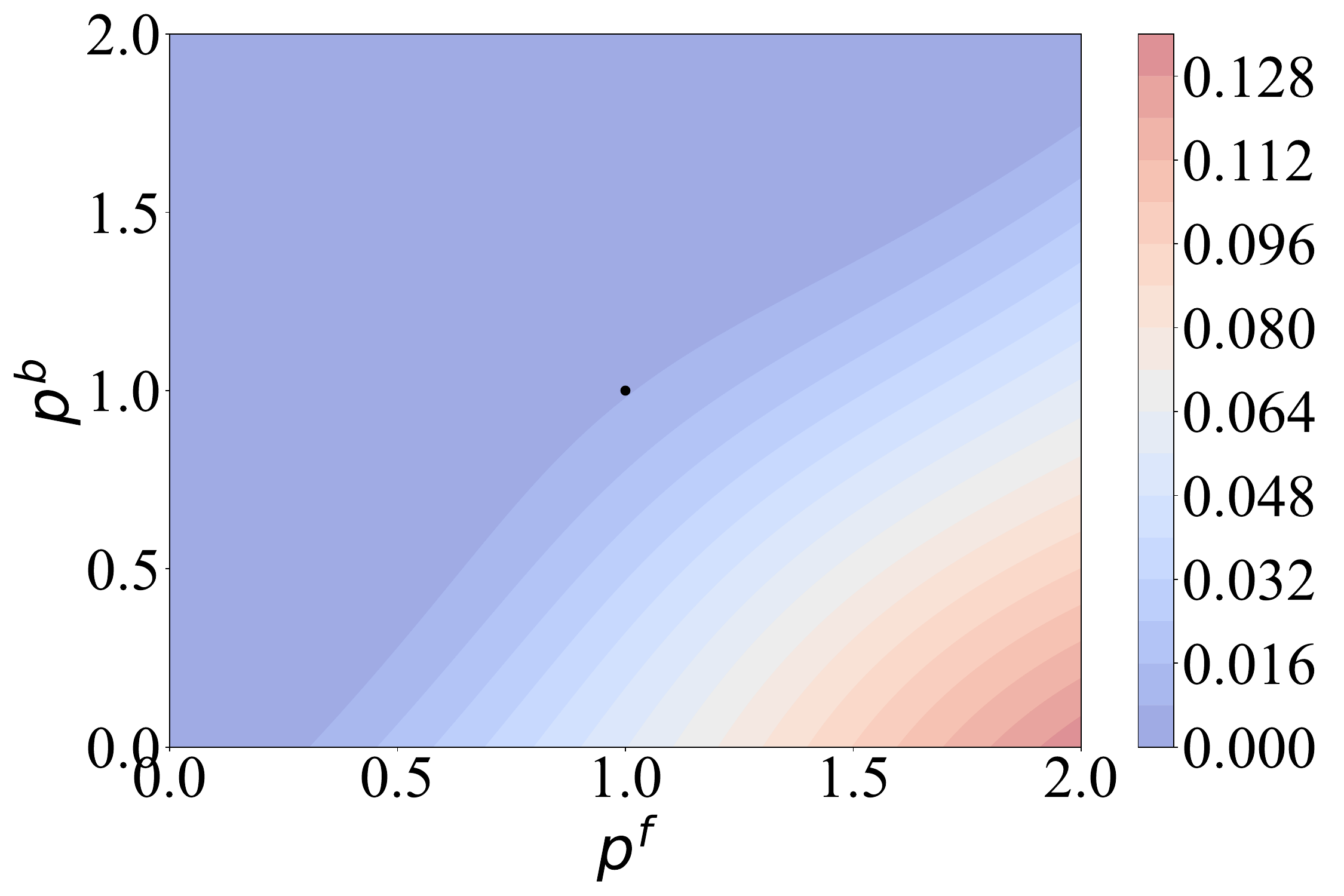}\label{fig:iss-contour-platoon1}}
    \subfigure[PlanarDrone2x2 ($p_l$ v.s. $p_r$)]{\includegraphics[width=.45\textwidth]{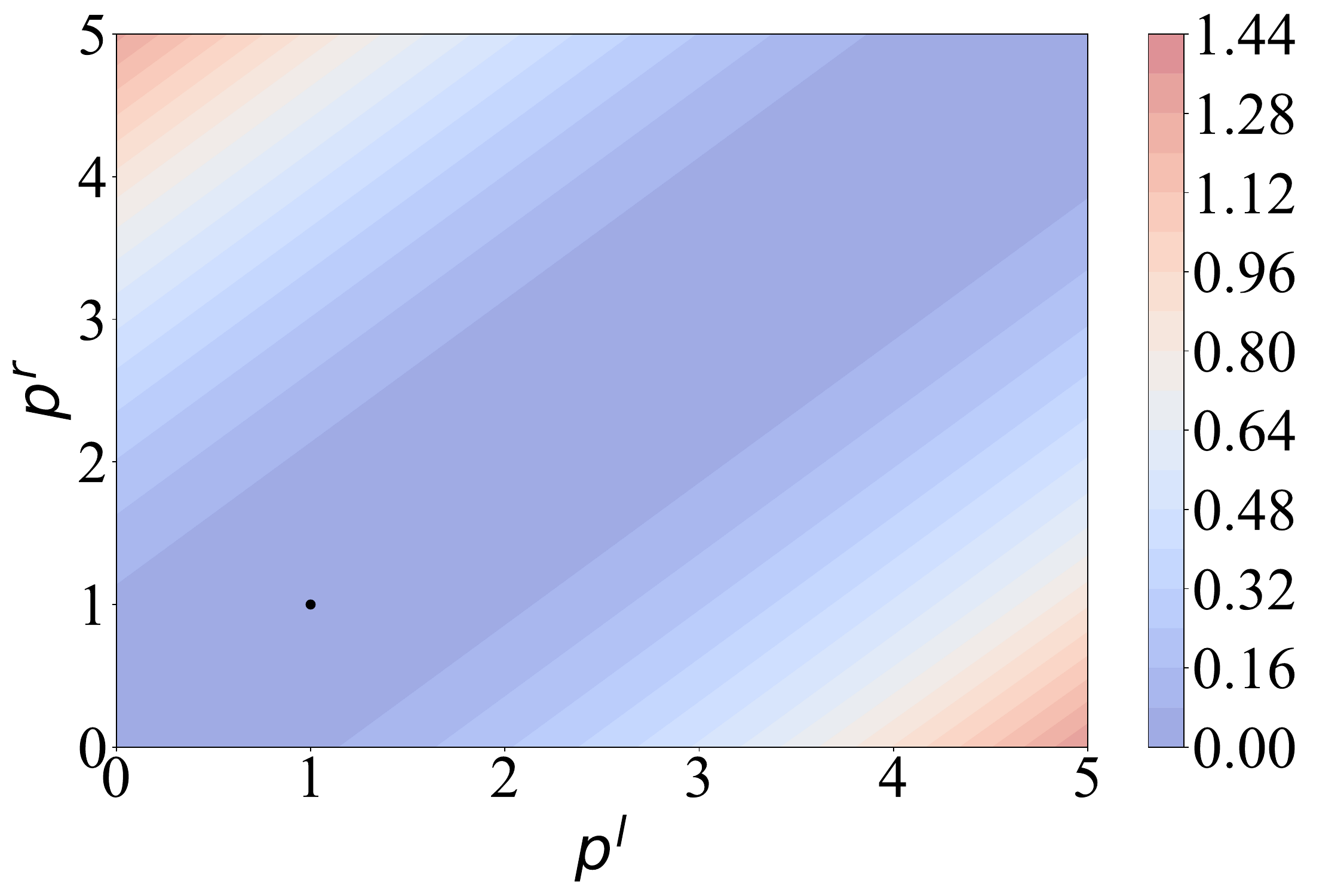}\label{fig:iss-contour-drone-lr}}
    \subfigure[PlanarDrone2x2 ($p_u$ v.s. $p_d$)]{\includegraphics[width=.45\textwidth]{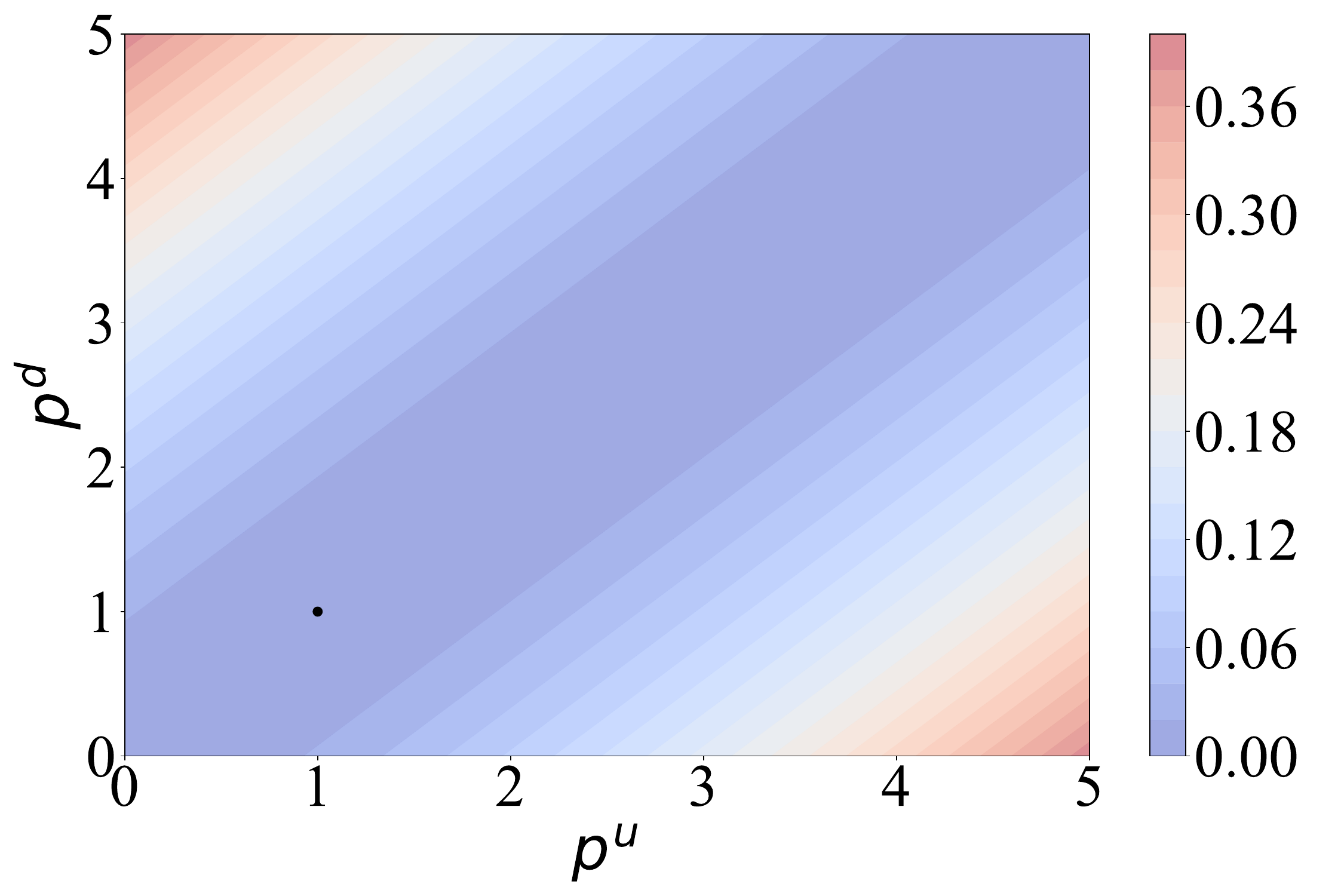}\label{fig:iss-contour-drone-ud}}
    \caption{Contour plots of the learned ISS Lyapunov functions, where the black dots show the goal point}
    \label{fig:iss-contour}
\end{figure*}

\subsubsection{Training Details}

In our framework, we include the hyper-parameters $\alpha_i$, $\epsilon_A$, $\epsilon_B$, $\mu_{A_i}$, $\mu_{B_i}$, $\mu_\mathrm{ctrl}$. For simplicity, we omit the coefficient of the first term $\mu_\mathrm{goal}$ in loss \eqref{eq:total-loss} in the main pages. We also include this hyper-parameter here. Note that omitting this coefficient in the main pages does not cause any problem because the coefficients $\mu_\mathrm{goal}$, $\mu_{A_i}$, $\mu_{B_i}$, $\mu_\mathrm{ctrl}$ only controls the weight of each loss. If we set $\mu_\mathrm{goal}=1$ manually and divide other coefficients and the learning rate with $\mu_\mathrm{goal}$, we can get the same results. The exact values of the hyper-parameters are included in \Cref{tab:hyperparam}.

We further discuss the function of each hyper-parameter. $\alpha_i$ is the convergence rate of the ISS Lyapunov function. Larger $\alpha_i$ can make the closed-loop system converge faster to the goal, but it also makes the training harder. $\epsilon_A$ and $\epsilon_B$ are used to encourage strict satisfactions of loss \eqref{eq:loss-cond} and \eqref{eq:loss-decent}, and to encourage the generalization abilities \citep{dawson2021charles}. Larger $\epsilon_A$ and $\epsilon_B$ make the learned ISS Lyapunov functions have better generalization abilities but also make the training harder. $\mu_\mathrm{goal}$, $\mu_{A_i}$, $\mu_{B_i}$ are weights of different terms in the total loss \eqref{eq:total-loss}. We choose them by balancing the value of each term to be at similar order of magnitudes. $\mu_\mathrm{ctrl}$ controls the strength of the additional training signal of $\pi_i(x_i;\theta_i)$. If the nominal controller is good, we can use large $\mu_\mathrm{ctrl}$ to accelerate the training, while if the nominal controller behaves badly, we use smaller $\mu_\mathrm{ctrl}$ or even set $\mu_\mathrm{ctrl}=0$ to make sure that the nominal controller will not affect $\pi_i(x_i;\theta_i)$ too much.

During training, we set the batch size to be $2048$, except for GridVoltage8 where the batch size is $1024$, and train \algo\ for $10000$ iterations. We set the learning rate to be $3\times10^{-4}$ for the ISS Lyapunov functions $V_i(x_i;S_i,\omega_i,\nu_i$, $5\times10^{-4}$ for the controllers $\pi(x_i;\theta_i)$, and $10^{-3}$ for the coefficients $k_i$ for $\chi_i(x_i;k_i)$. To prevent overfitting, we add the weight decay mechanism with coefficient $10^{-3}$ for the ISS Lyapunov functions and the controllers. We train \algo\ and the baselines $4$ times with random seeds $0,1,2,3$.

\subsubsection{Discussion on the GridVoltage8 Environment}

For GridVoltage8, PPO, MAPPO, and LYPPO have almost identically bad rewards and tracking errors. This is because we use a professional solver PandaPower \citep{pandapower} to simulate the underlying distribution grid. All three RL methods return controllers that quickly drive the system to an unsafe state, making the solver fail to solve the underlying model. In such cases, we simply set the reward and the tracking error assuming a 10\% voltage deviation, which is a typical safety limit for real-world distribution grids \citep{shi2022stability}.

\subsubsection{Additional Results}\label{sec:add_results}
In the main pages, we provide experimental results including comparison of the expected reward and the tracking error for small-scale environments, and the tracking error for large-scale environments. Here we provide more results of the experiments. 

We provide the expected reward of \algo\ and the baselines in the large-scale environments in \Cref{tab:reward-large}. Note that to avoid negative rewards, we change the reward function in PlanarDrone10x10 to be $r=100-\sum_i(|p_i^l-p_i^r|+|p_i^u-p_i^d|)$. The reward function in all the environments is used to measure the cumulative tracking error, which is used to compare the convergence speed of the tracking algorithms. We can observe that \algo\ achieves the highest reward in both environments, which means \algo\ converges the fastest.

We provide the contour plots of the learned ISS Lyapunov functions in \Cref{fig:iss-contour}. In the networked microgrid environments, we train only one ISS Lyapunov function for all the subsystems because of its robustness. The contour plot of the learned ISS Lyapunov function is shown in \Cref{fig:iss-contour-mg5}. We can observe that the learned ISS Lyapunov functions are in ellipse-like shapes indicating that the learned controller can go downhill w.r.t. the learned ISS Lyapunov functions and converge to the goal points (the black dots). In the networked GridVoltage8 environments, we train 8 ISS Lyapunov functions, one for each subsystem. The curve plots of the learned ISS Lyapunov function are shown in \Cref{fig:iss-contour-gv8}. Since we have only one input state for each subsystem, the learned ISS Lyapunov functions are in quadratic-like shapes, and the learned controller can go downwards w.r.t. the learned ISS Lyapunov functions and converge to the goal points (the black dots).
In the truck platoon environment, we train two ISS Lyapunov functions. One for the first and the last controllable truck ($i=1$ and $i=n$), and another one for all other trucks ($i=2,3,\ldots,n-1$). The contour plots of the learned ISS Lyapunov functions are shown in \Cref{fig:iss-contour-platoon0} and \Cref{fig:iss-contour-platoon1}. For trucks with $i=1,n$, the learned ISS Lyapunov functions will make them converge to the line where $p^f=p^b$. For trucks with $i=2,3,\ldots,n-1$, the learned ISS Lyapunov functions are sheer under the line $p^f=p^b$ and flat above that line. This is because the environment is parameterized by the velocity of the leading truck $v_0$, which is large compared with the initial velocity of other trucks. This causes that for the trucks in the middle, it is more often for them to reach the states where $p^f>p^b$, the region under $p^f=p^b$. Therefore, when the trucks are in states $p^f>p^b$, the sheer ISS Lyapunov functions will give them a strong signal to return to the line $p^f=p^b$. However when the trucks are in states $p^f<p^b$, they do not have to return to the $p^f=p^b$ quickly because the large $v_0$ can make them return automatically. In the drone formation control environment, we still train only one ISS Lyapunov function for all the subsystems because of its robustness. The learned ISS Lyapunov function is shown in \Cref{fig:iss-contour-drone-lr} and \Cref{fig:iss-contour-drone-ud}, which can make the drone converge to the states where $p^l=p^r$ and $p^u=p^d$. 

\end{document}